\documentclass[fleqn,usenatbib]{mnras}
\usepackage{tabularx}
\usepackage{booktabs}

\usepackage{siunitx}

\usepackage[T1]{fontenc}

\DeclareRobustCommand{\VAN}[3]{#2}
\let\VANthebibliography\thebibliography
\def\thebibliography{\DeclareRobustCommand{\VAN}[3]{##3}\VANthebibliography}

\usepackage{graphicx}	
\usepackage{amsmath}	
\usepackage{subfigure}
\usepackage{newtxtext,newtxmath}

\newcommand\ramses    {{\sc Ramses}}
\newcommand\MHL    {{\sc M6-Infall24}}
\newcommand\MHH    {{\sc M6-Infall23}}
\newcommand\MLL    {{\sc M5-Infall24}}
\newcommand\MLH    {{\sc M5-Infall23}}

\newcommand\DeltaYmax {\Delta Y_{\rm max}}
\newcommand\Msun {M_{\odot}}
\newcommand\Req {R_{\rm eq}}
\newcommand\sigpg {\sigma_{\rm pg}}
\newcommand\tinf {t_{\rm I}}
\newcommand\tSN {t_{\rm SN}}
\newcommand\vpg {v_{\rm pg}}

\title[Star formation in globular clusters]{Second generation star formation in globular clusters of different masses}

\author[A. Yaghoobi et al.]{
A. Yaghoobi$^{1,2}$\thanks{E-mail:a.yaghoobi@iasbs.ac.ir}, 
F. Calura$^{3}$,  
J. Rosdahl$^{2}$,  
H. Haghi$^{1}$
\\
$^{1}$Department of Physics, Institute for Advanced Studies in Basic Sciences (IASBS), 444 Prof. Yousef Sobouti Blvd., 45137-66731, Zanjan, Iran \\
$^{2}$ Univ Lyon, Univ Lyon1, Ens de Lyon, CNRS, Centre de Recherche Astrophysique de Lyon UMR5574, F-69230, Saint-Genis-Laval, France\\
$^{3}$INAF - OAS, Osservatorio di Astrofisica e Scienza dello Spazio di Bologna, via Gobetti 93/3, I-40129 Bologna, Italy
}

\date{Accepted XXX. Received YYY; in original form ZZZ}

\pubyear{2015}

\begin{document}
\label{firstpage}
\pagerange{\pageref{firstpage}--\pageref{lastpage}}
\maketitle
%
 
\begin{abstract}
  By means of three-dimensional hydrodynamical simulations, we
  investigate the formation of second generation (SG) stars in young
  globular clusters of different masses.  We consider clusters with a
  first generation of asymptotic giant branch (AGB) stars with mass
  $10^5$ and $10^6M_{\odot}$ moving at constant velocity through a
  uniform gas with density $10^{-24}$ and $10^{-23}$ g cm$^{-3}$.  Our
  setup is designed to reproduce the encounter of a young cluster with
  a reservoir of dense gas, e. g. during its orbital motion in the
  host galaxy.  In the low-density models, as a result of the cooling 
  AGB ejecta which collect in the centre, weakly perturbed by the external ram pressure, a compact
  central He-rich SG stellar component is formed on a timescale which
  decreases with increasing initial cluster mass. 
Our high-density models are subject to stronger ram pressure, which prevents the accumulation of
the most He-rich AGB ejecta in the cluster centre. 
As a result, the SG is more extended and less He-enhanced than in the low-density models.  
By combining our results with previous simulations, we are
  able to study relevant, cluster-related scaling relations
  across a dynamical range of two orders of magnitude in mass (from
  $10^5 M_{\odot}$ to $10^7 M_{\odot}$).  In agreement with current 
  observationally-based estimates, we find positive correlations between the SG-to-total
  number ratio and maximum He enhancement in SG stars as a function of the
  initial cluster mass.  
\end{abstract}

\begin{keywords}
Globular clusters: general – stars: formation – methods: numerical - hydrodynamics.
\end{keywords}



\section{Introduction}
In recent decades, many indications have been found that Galactic Globular Clusters (GCs) are not simple stellar populations (SSP), i. e. systems where all the stars share the same age and chemical composition, but that most of them host multiple stellar populations. The multiple populations (MPs) in a cluster differ in their abundance of light elements, with an ``anomalous'' component typically showing enhancements in He, N, and Na abundances, and depletions in C and O with respect to the field stars of the host galaxy \citep[e.g.][]{minniti1993,carretta2009,gratton2013, marino14}. However, in most cases the different populations are similar in their Fe content \citep{gratton2004}. 

In addition to the Milky Way (MW), these anomalous abundances have been observed in nearby galaxies such as the large and small Magellanic clouds \citep[hereafter LMC and SMC;][]{mucciarelli2009} Fornax \citep{larsen2014} and M31 \citep{nardiello2019}.
In the literature, the stellar populations within GCs with a ``normal'' chemical compositions, i. e. similar to ones of field stars, are usually referred to as first-generation (FG), first-population (1P) or primordial population. On the other hand, other populations with anomalous abundances are labeled as second-generation (SG), second-population (2P) or enriched population \citep{bastian2018}.
Through high precision HST photometry of red giant branch stars it has been revealed that the average number fraction of SG stars is 0.38 and 0.68 in GCs of the MW and the Magellanic clouds, respectively \citep[][]{milone2020}. 

Several scenarios have been proposed to explain the origin of multiple populations in GCs.
 At present, which scenario is favoured with respect to others is still a matter of debate \citep{renzini2015, bastian2018}. 
In most of these scenarios the chemical differences in FG and SG stars can be explained by a framework in which 
gas enriched by the ejecta of FG stars fuels the formation of SG stars. 

The various scenarios mainly differ in what are assumed to be the sources of the enriched gas. These include fast-rotating massive stars \citep{decressin2007}, massive binaries \citep{Mink2009,bastian2013},  supermassive stars \citep{denissenkov2013} and asymptotic giant branch \citep{D'Ercole2008} stars (AGBs). 

In the fast-rotating massive stars (FRMS) scenario, SG stars form from the enriched material ejected from massive stars (20-120 $M_{\odot}$) rotating with speeds near their break-up limit. Such material is still slow enough to be retained by the cluster and most of the SG star formation has to occur in a very short time, while the original cluster is younger than $3.5$ Myr  (\citealt{krause2013}), before the first supernovae explosions start to pollute the star-forming gas with Fe and eject most of the gas from the cluster, suppressing further star formation \citep{decressin2007}.

In the massive binary scenario the sources of enriched material are massive interacting binaries (10 - 100 $M_{\odot}$). These systems eject most of the envelopes of their primary stars into the interstellar medium (ISM) during non-conservative mass transfer and then SG stars can form from this enriched gas \citep{Mink2009}. The main advantage of this scenario is that it provides more enriched material than  the AGB and FRMS scenarios. 

Supermassive stars with mass $\approx 10^4 M_{\odot}$ and with significant mass loss have also been presented as a viable source for the origin of MPs \citep{denissenkov2013}. Due to  chemically homogeneous ejecta enriched in light elements, this model may account for the observed abundance anomalies in GCs. The problem is that such massive stars still only exist in the realm of theory, putting this scenario in significant doubt. 
 
An alternative scenario for SG formation was proposed by \citet{bastian2013}. According to this model, there is really only one generation of stars, and the reason for some low-mass FG stars having anomalous metal abundances is that they accrete or sweep up metal-enriched ejecta from massive binaries during their lifetime.

Another proposed scenario sees the presence of massive (with masses in the range $9-500 M_{\odot})$ metal-poor supergiants which induce
  an early star-formation episode, occurring within the first 4 Myr, i.e. before the explosion of the first core-collpse SNe \citep{szecsi2019}.
  This scenario predicts a positive correlation between the SG fraction and cluster mass and, as luminous supergiants are expected to be present only at low metallicity,
  it predicts a limited presence of a polluted second generation at high metallicity. 

The AGB scenario \citep{D'Ercole2008,D'Ercole2016}  is probably the one which has gained the most widespread attention and is the focus of this paper. Here, the SG of stars is formed out of AGB ejecta from the FG, which start to be ejected about $40$ Myr after the formation of the cluster, mixed with pristine\footnote{Throughout this paper, the pristine gas has the same helium abundance as FG stars, i. e. a helium mass fraction of Y=0.246.} gas accreted from the ambient ISM gas. 

By means of 1-D hydrodynamic simulations modelling this scenario, \cite{D'Ercole2008} showed that the SG stellar component is more concentrated than the FG stars and has enhanced He abundances. In this model, the pristine gas accreted by the cluster is a vital component in order to explain the observed relative fractions of FG to SG masses as well as their present-day abundance pattern. \citet{D'Ercole2011} showed that the large reservoir of pristine gas is also necessary to dilute the AGB ejecta and thus explain the correlations between light elements.

To form a massive SG cluster which incorporates a significant fraction of AGB ejecta, an initially very massive cluster is required, larger than its present-day mass typically by factors between 5 and 20 \citep{D'Ercole2008,renzini2015}. Hence within this scenario, it becomes very important to study the capability of a cluster to retain its AGB ejecta and to accrete pristine gas.

Moreover, this model has to cope with empirical evidence emerging from recent observations, which show a strong positive correlation between the number ratio of SG to FG stars and the present-day total mass of clusters hosting MPs \citep{milone2017,milone2020,baumgardt2018,bastian2018}. The inferred initial masses of such clusters appear to be between $10^5 M_{\odot}$ and $10^7 M_{\odot}$ \citep[e.g.][]{milone2009,mackey2008,baumgardt2019,carretta2010}.

From these pieces of evidence and the general expectation that more massive clusters accrete gas more easily, one can expect that the initial mass of the cluster plays a fundamental role in determining its ability to form multiple stellar  populations \citep{bastian2018}. Several studies have focused on the initial mass, modelling the main physical processes regulating star formation in a young cluster, such as its gravity \citep{pflamm2009}, gas accretion \citep{naiman2011}, ram pressure stripping \citep{charlie2010}, dynamical interactions between individual stars and the accreted gas \citep{bekki2019}, and stellar feedback \citep{naiman2018,D'Ercole2008,vesperini2010}. Some of these processes depend directly or indirectly on the mass of the star cluster and the properties of the external environment \citep{calura19, lin2007,charlie2010}, including gas density and temperature, and velocity of the cluster though the surrounding ISM. 

By means of one-dimensional hydrodynamical simulations, \citet{naiman2018} studied, in a range of cluster masses, the role played by other likely relevant parameters such as compactness, metallicity, and cluster age on the capability of isolated, spherically  symmetric  clusters to retain the ejecta of their stars. They found that isolated clusters need special conditions to retain mass: they essentially need to be compact and massive, with a velocity dispersion $\sigma>25$ km/s, corresponding to a cluster with a mass $10^7 \Msun$ and a core radius of 27 pc. 

Other studies have focused on a cluster moving through the ISM \citep[e.g.][]{lin2007,naiman2011}. These works confirm that if the central velocity dispersion of the cluster is greater than the sound speed of the ISM, as well as the relative velocity between the cluster and ISM, the cluster can accrete a significant amount of the ambient gas. However no direct prediction has been provided regarding the amount of gas which can be converted into new stars. 

Following up on previous results, recently \citet{calura19} (hereafter C19) presented the first set of 3-D hydrodynamic simulations, taking into account self-gravity and radiative cooling, aimed to study of the relative roles of stellar winds from AGB stars and the accretion of pristine gas in the formation of SG stars. 
The C19 simulations focus on the feasibility of the AGB scenario proposed by \citet{D'Ercole2016}, with a FG globular cluster of $10^7 M_{\odot}$ and a half-mass radius of $\approx 30$ pc.  This cluster was placed into a uniform distribution of $10^4$ K gas with two different densities, $10^{-23}$ and $10^{-24} \ {\rm g/cm^3}$, through which the cluster was moving with a speed of $20$ km/s.  In both these simulations, a compact SG sub-component was formed with a mass of $5\times10^6 M_{\odot}$ and $7\times10^5 M_{\odot}$ in the high-density and low-density cases, respectively. A distinct feature of their results was that the first and most helium-enhanced SG stars were born at the centre of the cluster, whereas stars born later and with lower He content were characterised by more extended spatial distributions. 

This predicted compactness of the SG stars is qualitatively in agreement with a few observational studies, in which clusters are found to retain some memory of their initial stellar distribution and the SG is found to be more spatially concentrated than the FG \citep[see e.g.][]{sollima2007,simioni2016}. 

Some of the standing problems of the AGB scenario require attention and are the subject of the present paper.  
To further explore the viability of the AGB scenario, the simulations from C19 now need to be expanded to a further, 
more detailed investigation of structural parameters that can also  be checked against observations. 
The aim of the present paper is to extend the study carried out by C19 to a range of initial cluster masses, allowing us to study the capability of less massive clusters to
retain their stellar winds, to accumulate pristine gas and transform them into stars. 

We start from a setup similar to that of C19 and study SG star formation in clusters of ten and a hundred times lower mass, $10^5$ and $10^6$ $M_{\odot}$.  By combining our results with those of C19, we  cover a dynamic range of three orders of magnitude and investigate the scaling of some fundamental properties like the SG mass and the fraction of AGB ejecta incorporated into new stars with initial cluster mass. Having a range of cluster masses also allows us to investigate whether the AGB model can reproduce the observed correlation between the number ratio of SG to FG stars and cluster mass.
 
The paper is organised as follows. In Section \ref{setup.sec} we describe our simulation setup and our most critical assumptions. In Section \ref{results.sec}, we present the results of our set of simulations. In Section \ref{discussion.sec} we discuss the most fundamental implications of our results and compare them with relevant observed properties of present-day GCs. Finally, in Section \ref{conclusions.sec} we present a summary and the main conclusions of our study.

\section{Simulation setup}\label{setup.sec}
The simulation setup used in this work is similar to that described in C19. We perform a series of 3-D simulations with the goal of studying the formation of a new generation of stars in a $\approx 30$ Myr old globular cluster, which roughly corresponds to the lifetime of the least massive type II SNe progenitors in the FG. This means that at the beginning of our simulations, the energetic feedback of massive FG stars has completely cleared out the residual, metal-rich gas polluted by the SNe (e. g., \citealt{calura2015}). 

The simulations are intended to study two critical parameters in the SG formation: the initial mass of the globular cluster (i.e. the FG mass) and the gas density of the surrounding ISM. Both these parameters  determine whether the cluster can accumulate the required mix of material to form a SG of stars. 

We use the \ramses{} code \citep{teyssier2002} to solve for the interactions of gas and SG stars via gravity, hydrodynamics, and radiative cooling on a uniform mesh. The gas advection is calculated by means of a second-order Godunov scheme to solve the Euler equations and the dynamical evolution of the SG stars is calculated with a Particle-Mesh solver. We use the acoustic Riemann solver for gas advection. We use an adiabatic index of  $\gamma = 5/3$ for the ratio between internal energy and gas pressure.  For simplicity, we neglect the dynamical evolution of FG stars, which are modelled with a static Plummer density profile \citep{plummer1911}.
In all our simulations, the size of the computational grid is $50$ pc which is uniformly divided into $512^3$ cells, corresponding to a spatial resolution of $0.1$ pc.  

Gas accretion and a subsequent formation of SG stars is only possible if we neglect feedback from massive SG stars (i.e. stars with mass $M > 8 M_{\odot}$ ).
Such feedback efficiently suppresses SG formation \citep[see][]{D'Ercole2008} and the supernova (SN) ejecta would produce a spread in heavy elements such as Fe,
which is found only in a small subset of the sample of Galactic clusters studied so far \citep{renzini2015}.

For SG stars, we therefore assume an initial mass function (IMF) truncated at $M \ge 8 M_{\odot}$.
 By necessity, the same assumption regarding the IMF has been made in previous theoretical works on the formation of SG stars \citep[e.g. C19, ][]{D'Ercole2008,D'Ercole2010}. 
The implication of this assumption is that stellar feedback from SG massive stars,
in the form of stellar winds, SN explosions and radiative feedback, is not present.
However, feedback from FG stars may still be present, e.g. in the form of ionising radiation. 
 An investigation of the effects of this energy source is currently in progress and will be presented in a forthcoming work.
A discussion of the realism of a truncated IMF for the SG will be presented in Sect. \ref{sec_IMF}. 
 
%
\begin{table*}
	\centering
	\begin{tabular}{lcccccc} 
		\hline
		Simulation & M$_{{\textup FG}}$ \ ${\rm [M_{\odot}]}$ & $\sigma$ \ $[10^6 \ {\rm cm \ s^{-1}}]$ & $\rho_{ \small pg}$ \ $[\rm g \ cm^{-3}]$ & $R_{\small eq}$ \ $[\rm pc]$ & $t_{\small I}$ \ $[\rm Myr]$ & $t_0$  \ $[\rm Myr]$ \\		\hline
		\MHL{}  & $10^6$ & 2.68&  $10^{-24}$ & 414&  43.5 & 12.2\\
		\MLL{} & $10^5$ & 0.85 & $10^{-24}$& 131& 34.3 & 3\\
		\MHH{} & $10^6$ & 2.68&  $10^{-23}$ & 131&  34.3 & 3\\
		\MLH{} & $10^5$ & 0.85 & $10^{-23}$& 41&  31.3 & 0\\
		\hline
	\end{tabular}
	\caption{Main simulation parameters. Here M$_{\rm FG}$ is the mass of the cluster, $\sigma$ is its velocity dispersion and $\rho_{\rm pg}$ is the ambient density. The stalling radius R$_{\rm eq}$ and the time at which the infall starts t$_{\rm I}$ are computed by means equations (\ref{eq:Req}) and (\ref{eq:tI}), respectively. $t_0$ is the infall time in the time reference frame assumed in this study (i.e. $t_I-31.3$ Myr).}
	\label{tabl1}
\end{table*}

\subsection{Initial conditions}
Following \citet{D'Ercole2016}, the cluster is assumed to belong to a high-redshift disky dwarf galaxy. Most globular clusters are indeed believed to be relics from the high-redshift Universe \citep{kravtsov2005,kruijssen2015}. 

After the rapid formation of FG stars, the most massive stars explode as supernovae (SNe), blowing the local gas out of the cluster, suppressing further star formation and powering a hot gas bubble  \citep{weaver1977,calura2015,gavagnin2017}. Supported by continuous SN II explosions, the bubble expands, sweeps up the ambient gas and blows out of the disk, dispersing the metal-rich bubble material into the circum-galactic medium.  A galactic disk is an ideal environment for our scenario because of the large reservoir of gas that will be available for later accretion and because its limited vertical extent allows the blowout of the bubble, preventing the incorporation of SN II ejecta into SG stars.

The cluster can thus accrete mass from the galactic ISM while orbiting around the centre of the galaxy \citep[see, e. g.][]{goodman2018}. In principle, the cluster may be located in a gas distribution with a non-disky geometry which can represent other real cases such as a reservoir of gas left over from a merger or simply an irregular galaxy. At this stage, the shell confining  the bubble is suddenly accelerated, and becomes prone to Rayleigh-Taylor instability which may lead it to disruption. At the break-out, the hot interior of the bubble leaks out and its pressure drops. The bubble loses its spherical shape and the wind due to continuous SN explosions impinges directly on the inner part of the shell. When SN explosions cease, the further expansion of the shell can continue, driven by the inertia of the swept-up mass.

The bubble stalls as soon as the expansion velocity equals the velocity dispersion of the unperturbed ISM, which is of the order of the local sound speed \citep{D'Ercole2016}. At this stage, the shell merges with the external ISM, the bubble loses its initial structure and can be regarded as a hole in the disk carved by the feedback of massive FG stars.  The cavity then starts to be re-filled by ISM gas pouring in at a velocity on the order of the local sound speed. 

The stalling radius of the bubble can be computed as (C19, \citealt{ D'Ercole2016}):

\begin{align}\label{eq:Req}
\Req=4.143 \times 10^3 \ {\rm pc } \ \ \left( \frac{L_{41}}{n_0 V_{w,8} \ (\sigma_{\rm pg,6}^2+v_{\rm pg,6}^2)}\right)^{1/2}.
\end{align}
Here $L_{41}$ is the mechanical luminosity of FG SNe  in units of $10^{41}$ ergs $s^{-1}$, $V_{w,8}$ is the velocity of the SN ejecta in units of $10^8 \ {\rm cm \ s^{-1}}$, $\sigma_{\rm pg,6}$ and $v_{\rm pg,6}$ are respectively the isothermal sound speed and the velocity of the pristine gas relative to the globular cluster (both in units of $10^6$ cm/s), and  $n_0$ is the ISM  number density in $\rm cm^{-3}$. We assume that at a time ${\rm t_{I}}$ after its birth, the cluster reaches the edge of the feedback bubble as described above and starts moving through the unperturbed ISM. 

We refer to this moment as the infall time $\tinf{}$, defined as
\begin{equation}\label{eq:tI}
\tinf=\tSN+\frac{\Req}{\sigpg+\vpg},
\end{equation}
where $\tSN\approx30$ Myr is the time at which we assume SN explosions to cease. 

Our aim is to mimic the motion of the cluster into the cavity carved by FG massive stars and a consequently asymmetric accretion of gas.
As discussed in \cite{D'Ercole2016}, the accreted gas has not been enriched by FG core-collapse SNe. 
In fact, numerical simulations of nuclear starbursts have shown that 
the wind hitting the shell on the disc is deflected away from the plane, ablating the 
superficial gas layers of the inner side of the shell and dragging 
them away (\citealt{tenoriotagle1998}). Moreover, due to numerical diffusion, the mixing achieved in numerical simulations 
is stronger than in reality. In another model, \cite{tenoriotagle1996} have shown that, even if 
 the metal-enriched blown-out gas mixes efficiently in a hot 
 gaseous halo and is later reaccreted by the system, this does not occur before $\sim 10^8$ yr.
 It is therefore reasonable to assume that, for the time interval of interest in our model,
 the GC wind does not contaminate the composition of the host galaxy. 

In our reference frame the cluster is at rest and at time $\tinf$ the infalling pristine gas enters the box on one side with a velocity $\vpg$ parallel to the x-axis.  
The chemical composition of the pristine gas is assumed to be the same as the that of the FG stars, with a He mass fraction Y=0.246 and a metal mass fraction Z=0.001. As in C19, we assume $\sigpg= 1.16 \times 10^6$ cm/s and $\vpg=2.3 \times 10^6$ cm/s.
 As current observations indicate a flat relation between the mass and the half-mass radius for young massive clusters and GCs \citep{portegies2010,ryon2017,krumholz2019} with mass $<10^6~M_{\odot}$, 
for all our clusters we assume a half mass radius of 4 pc, corresponding to a Plummer radius $r_{\rm P} = 3$ pc. The main model parameters are listed in Table \ref{tabl1}.

\subsection{Star formation}\label{SF}
The star formation sub-grid model used in our simulation is described in \citet{rasera2006} and in C19. In this scheme, gas cells are eligible for star formation if the gas temperature $T<2\times10^4$ K and the gas flow is converging, i.e. $\nabla \cdot\mbox{\boldmath{$v$}} < 0$, where $\mbox{\boldmath{$v$}}$ is the velocity of the gas.
The gas in eligible cells is converted into star particles with an average rate per unit time expressed by a \citet[][]{schmidt1959} law
\begin{equation}\label{eq10}
   \dot \rho_{*}= \frac{\rho}{t_*},
\end{equation}
where $t_*$ is the star formation timescale. We assume a uniform star formation timescale of $t_*=100$ Myr (C19). 
Our results do not depend significantly on the choice of this parameter, as shown in \citet{D'Ercole2008}.

In each time-step, the gas may be converted into stellar particles, each having a mass of an integer multiple of $m_*=0.1 \ \Msun$, by sampling the Poisson probability distribution for conversion of the gas to stars via eq. \ref{eq10}.  We allow no more than 90\% of the cell gas to be turned into stars. This condition implies a minimum density threshold for star formation
\begin{equation}
\rho_{{\rm th}}=\frac{m_*}{0.9 \ (\Delta x)^3}=7.6 \times 10^{-21} \ {\rm g \ cm^{-3}}, 
\end{equation}
where $\Delta x=0.1$ pc is the cell width. 
The new star particle formed in a cell is placed at its centre and it is given a velocity equal to the cell gas.

At the beginning of the simulation, we assume the FG to be already in place.
The gravitational effect of the FG is modelled by means of an analytic potential. 
In our 'wind tunnel' setup, we use a reference frame in which the cluster
is at rest and is accreting mass from one of the simulation boundaries. 
FG stars are continuously distributed in space, therefore occupying all the cells, 
with a spatial distribution 
described by an analytic Plummer mass density profile: 
\begin{equation} 
\rho_{*,\rm FG}(r) = \frac{3 \ M_{\rm tot}}{4\pi\, a^3} \left(1+\frac{r^2}{a^2}\right)^{-\frac{5}{2}}, 
\label{plum}
\end{equation}
where $r$ is the radius from the centre of the cluster and $a=3$ pc. For the total mass of the FG cluster, $M_{\rm tot}$, we assume two different values, i.e. $10^5 \ M_{\odot}$ and $10^6 \ M_{\odot}$, in two sets of simulations. 

Our simulations are stopped 100 Myr after the formation of the cluster. This time corresponds to the onset of FG type Ia SNe, whose energetic feedback is assumed to halt star formation \citep{D'Ercole2008, maoz2014}.

\subsection{Heating and cooling}

FG AGB stars act as a heating source in our simulations by injecting mass and energy into their surrounding medium.
In each simulation cell, the mass injected by AGB stars per unit time and volume is
\begin{equation}\label{AGB_ej}
\dot{\rho}_{ \rm AGB}=\alpha \rho_{*,\rm FG}, 
\end{equation} 
where $\rho_{*,\rm FG}$ is the local FG density and 
\begin{equation}
    \alpha(\tau)=0.065 ~\tau^{-1.01} 
\end{equation}
is the specific mass return rate (in yr$^{-1}$) as a function of the age $\tau$ (expressed in yr) of the FG population computed for a \cite{kroupa2001} IMF. Following C19, the rate of energy injection per unit volume from the FG AGB stars is  
\begin{equation}\label{eq14}
S=0.5 \ \alpha \ \rho_{*,\rm FG} \ (3\sigma^2+v^2+v_{\rm wind}^2),
\end{equation}
where $\sigma$ is the 1D velocity dispersion of the cluster, whereas  $v_{\rm wind}$ and $v$ are the wind velocity of AGB stars and the velocity of the gas, respectively. We assume $v_{\rm wind} =2\times 10^6 \ {\rm cm \ s}^{-1}$ \citep[see][]{D'Ercole2008}.
 
We also include radiative cooling in its native implementation in the \ramses{} code due to hydrogen, helium, and metals  \citep[see][]{few2014}. We apply a temperature floor of  $10^3 K$. The inflowing ISM gas, as well as all the gas in the initial conditions, has a temperature of $10^4$ K, which is typical for the warm photoionised ISM \citep[e.g.][]{haffner2009}.

Regarding the chemical compositions of SG stars, in our simulations we focus only on the evolution of their helium (He) abundance and do not track the abundances of the other elements. To track the gas He abundance, we use a passive scalar which is advected with the gas, and we likewise attach a helium mass fraction to each stellar particle.

As in C19, the helium abundances used in our work
are from \cite{ventura2011}. For the 
helium yield of the 8 $M_{\odot}$ progenitor we adopt a value approximately equal to the average of the yields of the model
of \cite{ventura2011}  and by \cite{siess2010}, calculated for a metallicity Z = 0.001. 
This allows us to achieve a helium mass fraction in the AGB ejecta which varies between Y=0.36 at the beginning of the AGB phase (39 Myr) and Y=0.32,
at the assumed end of star formation (i.e. 100 Myr, \citealt{D'Ercole2008, D'Ercole2010}).

\section{Results}\label{results.sec}
\begin{figure*}
\centering
\includegraphics[width=\linewidth]{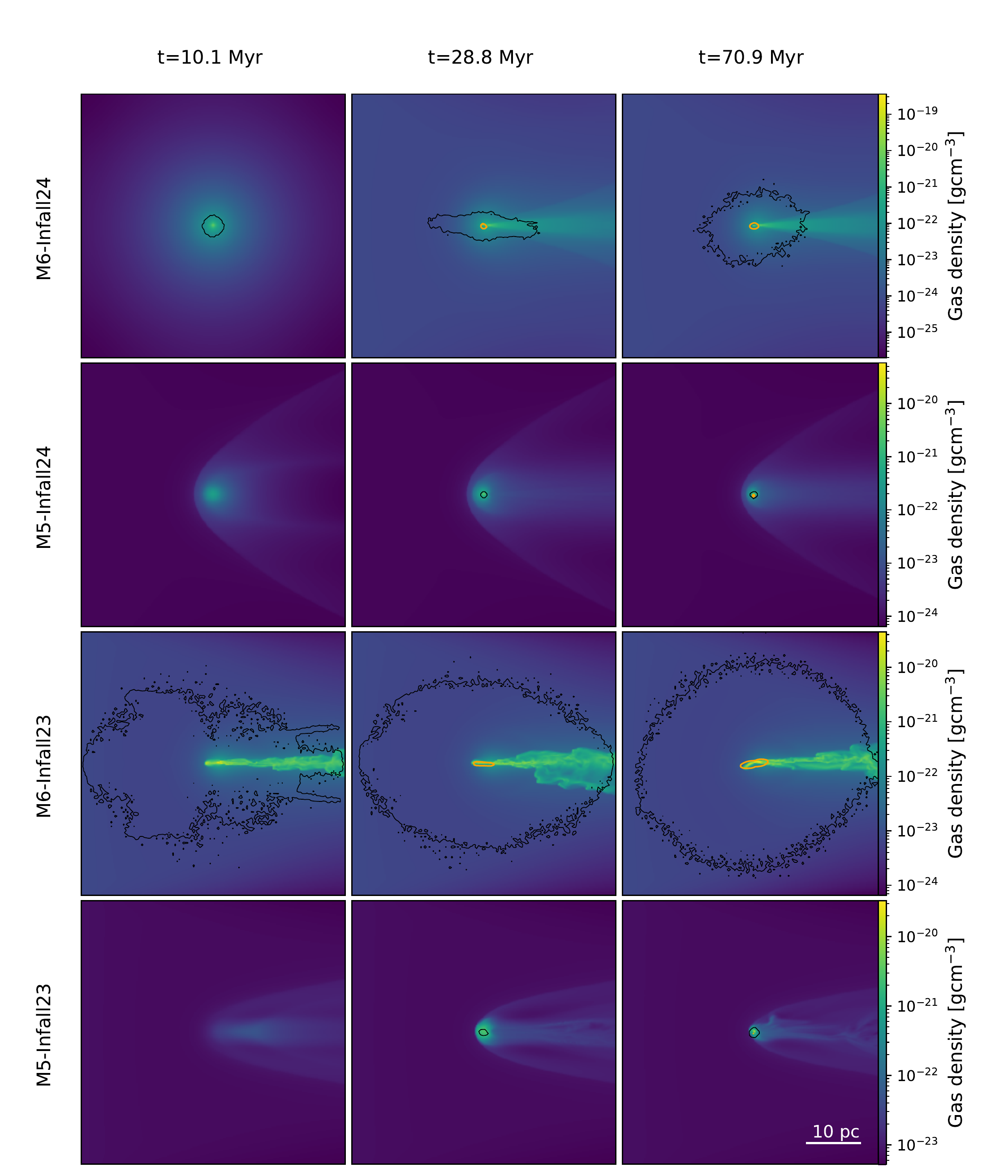}
\caption{Density slices computed in the x-y plane for our simulations, from top to bottom: \MHL{}, \MLL{}, 
  \MHH{}, \MLH{}) at three different times, as indicated above of each column.
  In each panel the black and orange contours enclose regions within which
the SG stellar density is $>6 \times 10^{-6}$ and $>0.06$ times the maximum value, respectively.  
    The physical scale is shown in the bottom-right panel. }
\label{fig:dens}
\end{figure*}

\begin{figure*}
\centering
\includegraphics[width=\linewidth]{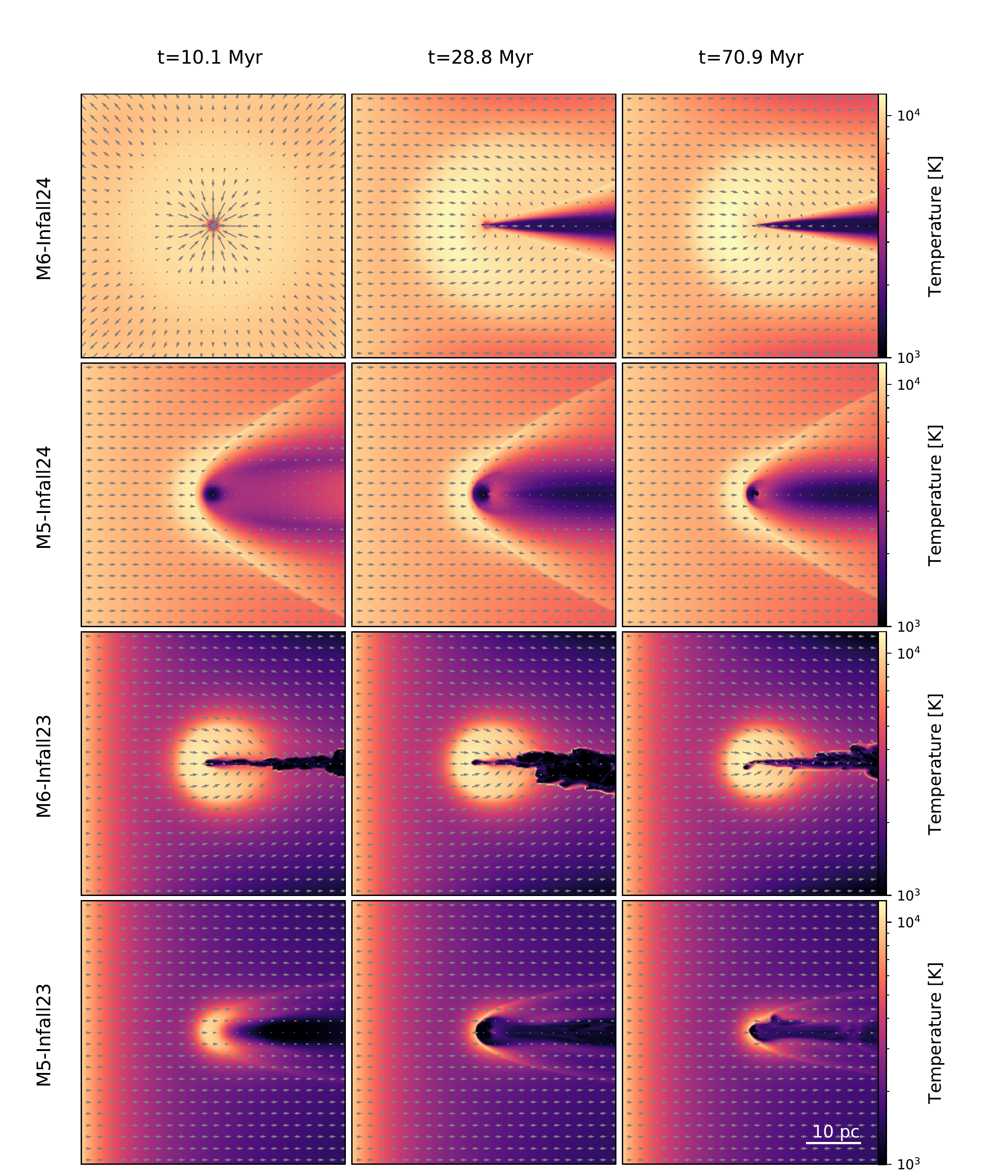}
\caption{Temperature slices computed in the x-y plane for our simulations, as indicated on the left, at three different times, as indicated on top. Gray arrows represent the gas velocity field.} 
\label{fig:temp}
\end{figure*}
\begin{figure*}
\centering
\includegraphics[width=\linewidth]{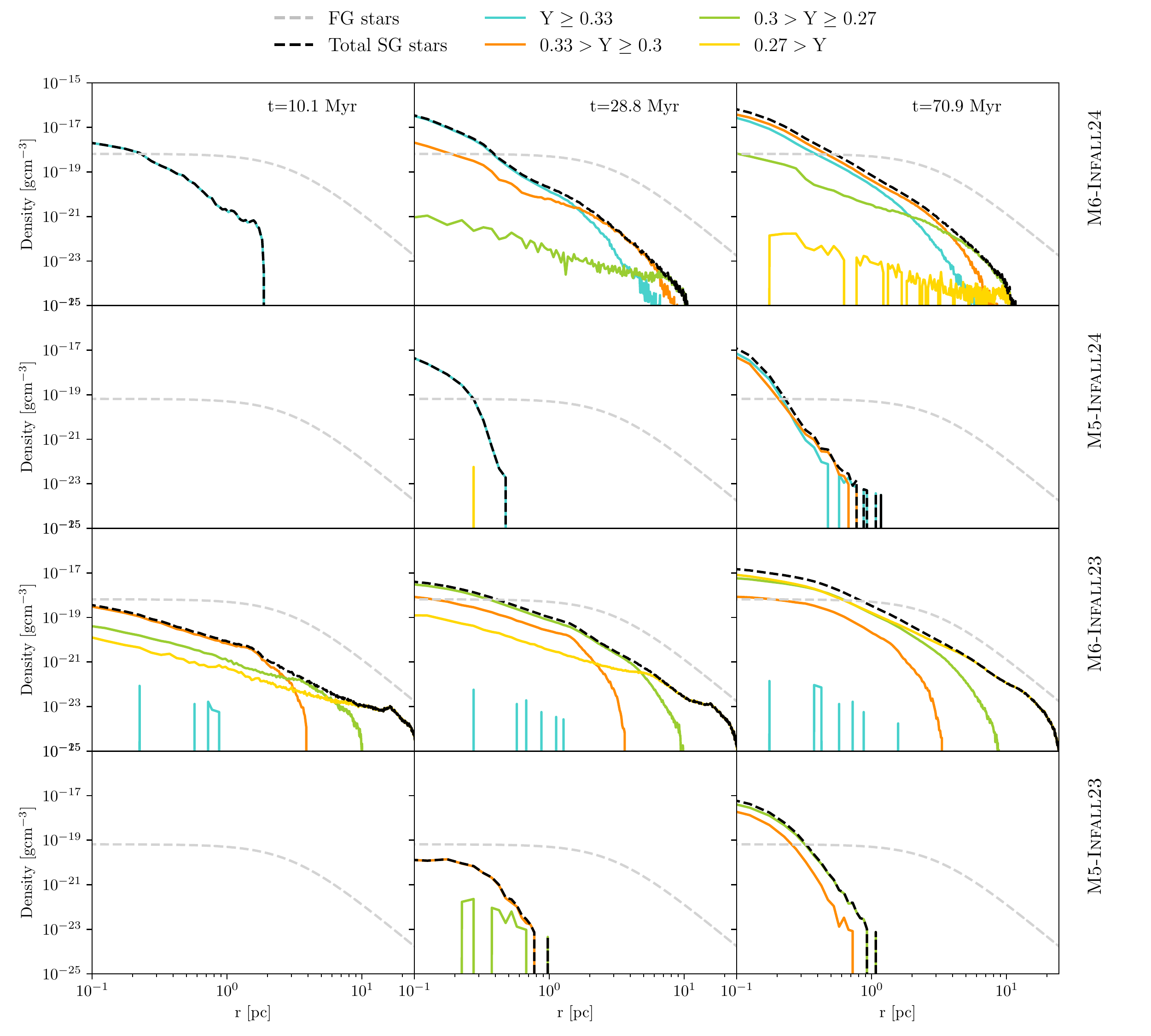}
\caption{First and SG stellar density profiles for various sub-components with different He abundance (see the legend in the bottom-left panel) computed at different times for our simulations. The times are indicated at the top and the names of the simulations are shown at the right of the figure. } 
\label{fig:profil}
\end{figure*}
\begin{figure*}
\centering
\includegraphics[width=0.7\linewidth]{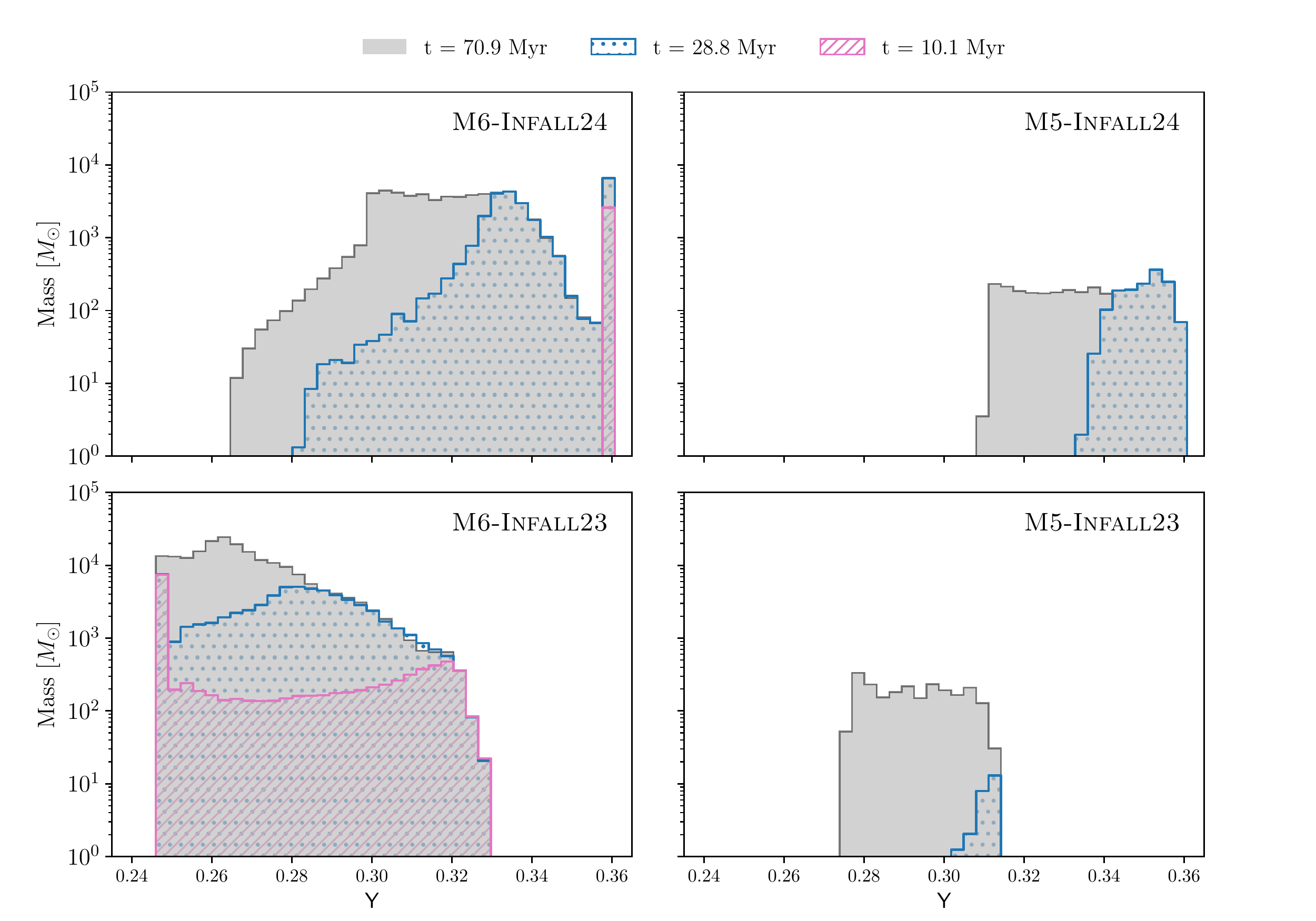}
\caption{Mass distribution of SG stars as a function of the He abundance $Y$ in SG stars at various times (see the legend in the top-right panel). From top-left, clockwise: results for the \MHL{}, \MLL{}, \MLH{}, and \MHH{}. 
The helium mass fraction of SG stars varies between Y=0.25, which is the helium mass fraction of the pristine ISM gas (and FG stars), and Y=0.36, originating purely from the most massive AGB ejecta at t=39 Myr.}
\label{fig:dist}
\end{figure*}

We now describe the results of our simulations, which again are listed in table \ref{tabl1}. We examine the effect of the initial cluster mass on the formation of SG stars and explore how and under what conditions ram pressure limits the ability of the gravitational potential to retain the stellar winds ejected by AGB stars and accumulate pristine gas. 

We expect that a cluster can retain its AGB ejecta only if it has a velocity dispersion higher than the velocity of the AGB ejecta, which may not always be true, especially at low mass \citep{naiman2018}. Similarly, in the case of a cluster moving in a gaseous medium, only a sufficiently massive cluster will be able to accrete enough mass to form new stars. This will happen only when the central velocity dispersion of the cluster is greater than the sum of the sound speed  and the relative velocity with respect to the external ISM  \citep{naiman2011,lin2007}. Based on these approximate analytical arguments, it is likely, as concluded in C19, that only clusters with mass larger than $10^6 M_{\odot}$ have the conditions to produce a new stellar generation. However, factors such as gas self-gravity, radiative cooling, and star formation are not properly accounted for in the analytic model, and hence it becomes relevant to study numerically even the case of lower-mass clusters. 

As already stressed, $t_{\rm I}$, i.e. the time at which we start the inflow of gas into the side of the simulation volume, varies with the FG cluster mass and the density of the unperturbed ISM. The minimum such time is $t_{\rm I}=31.3$ Myr. Thus throughout this paper, times are expressed by $t_0$ as the time after this minimum $t_{\rm I}$, i.e. $t_0=\tinf-31.3$ Myr. The values of both $t_{\rm I}$ and $t_0$ are given in the last two columns of Table \ref{tabl1}. Also, in this time reference frame the injection of AGB ejecta starts at $t_{\rm AGB}=7.7$ Myr in all simulations.

\subsection{Low-density simulations}
C19 showed that if a cluster with the a mass $10^7 M_{\odot}$ and  velocity $v_{pg}=23$ km/s moves in a medium with a density $10^{-24} g/cm^3$, a compact, massive SG with both He-rich and He-intermediate stars can form within the cluster. This was found to be in good agreement with observational data from very massive GCs. Here we study the same ISM density, but lower FG cluster masses of $10^6$ and $10^5 M_{\odot}$.

\subsubsection{Model \MHL{}}\label{M6-Infall24}
This model starts  with the injection of mass and energy from AGB stars at $t_{\rm AGB}=7.7$ Myr, followed 4.5 Myr later by the infall of pristine gas. 

The top row of Fig.\ref{fig:dens} shows gas and SG stellar densities from this simulation in the x-y plane at $10$ Myr, $29$ Myr and $71$ Myr. The orange and black contours indicate regions within which the 
SG stellar density is $>6 \times 10^{-6}$ and $>0.06$ times the maximum central value, respectively. 
The aim of the contours is to visualise the extent of all SG stars and a dense, central region where the stellar density is very high (typically about $10^{-17} \ {\rm g \ cm}^{-3}$ or $10^5 \Msun \ {\rm pc}^{-3}$). The top row of Fig. \ref{fig:temp} shows the gas temperature and velocity, computed at the same times. 

At $10$ Myr, AGB stars have already started ejecting enriched gas. The cluster can retain gas at radii $r$ where the cluster velocity dispersion $\sigma(r)$ > v$_{\rm wind}$. Ignoring the self-gravity of the gas and SG stars, the gravity only consists of the static Plummer potential.
It can be shown that this condition holds at radii $<20$ pc. This rough analytic estimate nicely matches the results shown in the top left density map in Fig. \ref{fig:temp}. The gas within this radius flows inwards and cools increasingly as the density increases. The cooling flow directed toward the cluster core is visible from the gray arrows in the top-left panel of Fig. \ref{fig:temp}, showing that the center of the cluster is occupied by cold and dense gas with temperature close to the temperature floor of $10^3$ K. This converging cold gas now comes under the correct conditions to form a SG of He-rich stars, and indeed a nearly spherical stellar distribution of SG stars is visible at the centre of the box, represented by the black contour in the top-left panel of Fig.\ref{fig:dens}.

At  $12.2$ Myr, an infall of pristine gas with velocity $v_{pg}=23$ km/s enters the box from the left, moving along the x-direction to the right. The high-velocity gas crosses the centre of the cluster, creating a wide cone-shaped tail downstream if it. Now the accretion of gas becomes regulated by the balance between the gravitational pull of the cluster and the ram pressure stripping of the incoming gas. Since in this simulation $\sigma^2>{V^2_{p g}+c^2_s}$ \citep{lin2007}, the cluster potential overcomes the ram pressure and, beside retaining AGB ejecta, the cluster begins to accrete pristine gas. 

At $29$ Myr a dense, symmetric cold tail is visible downstream of the cluster centre (top middle panel in Figs. \ref{fig:dens} and \ref{fig:temp}).  The arrows in the temperature map show the motion of the incoming gas. At the far left side of the box, the gas flows strictly towards the right, but as it flows further in, the gas is increasingly pulled towards the center and to the cold downstream tail. The tail itself flows in the opposite direction to the incoming ISM gas, towards the centre of the cluster from the right, in an ``accretion column'' \citep{bondi1944, shima1985}.

It is mostly thanks to the dense accretion column that pristine gas flows on towards the centre and mixes with the AGB ejecta.
Stars are forming along the column. Once formed, new stars are gravitationally bound to the center. In their orbits, some of the newly formed stars cross the center and may end up upstream of the cluster, moving in an opposite direction with respect to the gas flow. The elongated distribution of the stars reflects their motion as well as the dynamical state and extent of the gas out of which they form (see the black contour in the top-middle panel of Fig. ~\ref{fig:dens}).

At the final time ($t=71$ Myr), the accretion of pristine gas and confinement of AGB ejecta has allowed the stellar mass to grow considerably.  The shape of the stellar distribution has become more circular and diffuse (top-right panel of Fig. \ref{fig:dens}). 

Fig \ref{fig:profil} shows density profiles of FG and SG stars, in addition to sub-components of SG stars with different He contents. The density profiles are plotted at the same times as in Figures \ref{fig:dens} and \ref{fig:temp}, and are computed from the center of FG cluster, using concentric shells around the center and evaluating the stellar mass contained in each shell. The top row of the figure shows that early in the \MHL{} simulation, all the newly born SG stars are He-rich ($Y>0.33$), being fuelled almost entirely by AGB ejecta. The SG stars are located at the centre of the cluster, within the innermost $\approx 2$ pc. It is worth noting that SG stars already dominate the mass distribution in the cluster core. 

At a later time ($29$ Myr, top middle panel of of Fig. \ref{fig:profil}), the infall of pristine gas has started and the infalling matter mixed with the AGB ejecta is starting to be incorporated into new stars. Although the SG profile is still fuelled predominantly by AGB ejecta, we start to see the formation of stars with He abundances between the FG ejecta ($Y>0.32$) and the ISM ($Y=0.246$), which we refer to as He-intermediate stars, and which in the beginning have a less concentrated profile than the more He-rich SG stars. At $71$ Myr, an additional sub-population with $Y < 0.27$ is starting to emerge out of pristine gas (top-right panel of Fig. \ref{fig:profil}). The SG density profiles show that with time their distribution becomes increasingly extended and in general, the lower the helium abundance, the more extended the stellar distribution.

Fig. \ref{fig:dist} shows the SG mass-distribution of helium abundance. 
For \MHL{} (top left panel), a narrow, single-peaked distribution is seen at $\sim 10$ Myr, reflecting the He-rich population fuelled by AGB ejecta. At later times, as a result of the dilution of the AGB ejecta, the distribution becomes bimodal, with one ``mixed'' peak at $Y \approx 0.34$ and an ``enriched'' one at  $ Y \approx 0.36$. The distribution broadens and the SG stars gradually become less He-enriched towards the final time of $71$ Myr.

Eventually in this simulation, after $71$ Myr a compact SG is present, with a total mass of $6\times 10^4 \ \Msun$, which gives a SG-to-FG mass ratio of about 0.06, and a half-mass radius of 0.33 pc. 

\subsubsection{Model \MLL{}} 
We now report on a cluster with a ten times less massive FG ($10^5 M_{\odot}$). In this case, a smaller pre-simulation feedback bubble is expected and the infall of pristine gas starts earlier than in the case of the more massive FG from the previous sub-section.  By means of equations (\ref{eq:Req}) and (\ref{eq:tI}), we derive that the infall begins at $t_0=3$ Myr, i. e. 4.7 Myr before the onset of AGB ejecta at $t_{\rm AGB}=7.7$ Myr. 

In this case, the gravitational potential well of the cluster is weaker than in the previous simulation. Hence in the earliest phases, the ram pressure prevents the cluster from retaining a significant amount of gas.

As we can see from second-row panels of Fig. \ref{fig:dens}, at $t=10$ Myr a density enhancement is visible in the central region. The accumulated material is composed of a mix of pristine gas and AGB ejecta and is cooled down to $T<10^4$ K (second row of Fig. \ref{fig:temp}). Despite the central over-density, in this case no SG stars have formed yet. This is because the gas density has not reached the density threshold for star formation, i. e.  $7.8\times 10^{-21} \ {\rm g \ cm}^{-3}$ (see section \ref{SF}). At $t=29$ Myr, a very compact SG component is in place, visible from the black contour at the centre of the cluster. This component is mostly composed of He-rich stars ($Y \sim 0.35$, Figs. \ref{fig:profil} and \ref{fig:dist}),
i.e. mostly made out of AGB ejecta. The infalling gas is only weakly affected by the presence of the cluster and as a result no significant tail has formed at this time.

At 71 Myr, we see the appearance of a stellar component with an intermediate He abundance (second row right column panel of Fig. \ref{fig:profil}), due to dilution of the AGB ejecta. The final $Y$ distribution is much narrower than for the more massive FG cluster and in the range $ Y\sim 0.31-0.36$ (top right panel of Fig. \ref{fig:dist}).

The final SG stellar mass here is $ 3.2\times 10^3M_\odot$, corresponding to a FG-SG mass ratio of $0.03$, which is half of the ratio in the case of the more massive FG cluster from the previous sub-section.  The final half-mass radius is 0.08 pc. 

In table \ref{tabl2} we report the main properties of the SG population, computed at the end of our simulations. This includes the final SG mass, the SG/FG ratio, the fraction of AGB ejecta and pristine gas incorporated in SG stars, the He abundance pattern of SG stars (as indicated by the minimum and maximum values of Y) and half-mass radius of SG stars.

\begin{table*}
	\centering
	\label{tbl2}
	\begin{tabular}{lcccccccr} 
		\hline
                Model & FG mass \ [$M_{\odot}$] &  SG mass \ [$M_{\odot}$] & $f_{ \rm SG/FG}$ & $f_{\rm AGB}$  &$f_{\rm P}$  &$Y_{\rm min}$   &$Y_{\rm max}$  & $r_{\rm h,SG}$ [pc]\\
		\hline
		\MHL{} & $10^6$       & 6.3$\times10^4$ &0.063  &0.77 &0.23   &0.264 &0.36  & 0.33  \\
		\MLL{} & $10^5$       & 3.2$\times10^3$ &0.032  &0.91 &0.09   &0.31  &0.36  & 0.08 \\
		\MHH{} & $10^6$       & 2.0$\times10^5$ &0.205  &0.23 &0.77   &0.246 &0.331  & 1.52  \\
		\MLH{} & $10^5$       & 2.3$\times10^3$ &0.023  &0.49 &0.51   &0.273 &0.314  & 0.14  \\
      		\hline
	\end{tabular}
        
	\caption{Main results obtained at the end of each simulation. From left to right, the columns show: the name of the model, the mass of the FG cluster, the final SG mass, the final SG mass fraction ($f_{ \rm SG/FG}$), the fractions of AGB ejecta and pristine gas incorporated in SG stars ($f_{\rm AGB}$ and $f_{\rm P}$ respectively), the minimum and maximum He of SG stars ($Y_{\rm min}$ and $Y_{\rm max}$) and the half-mass radius of SG stars $r_{\rm h,SG}$.}
	\label{tabl2}
\end{table*}
\subsection{High-density simulations}
A higher ISM density than the value $\rho=10^{-24} {\rm g \ cm}^{-3}$ explored in the previous sub-section is expected to have a strong impact on the capability of the cluster to accumulate mass \citep{naiman2011} and to form SG stars. According to C19, when a very massive cluster moves through a denser medium, a more massive and extended -- but also less He-enriched -- SG can form, due to a larger amount of accreted pristine gas. In this section we thus study simulated scenarios with the same two FG masses as in the previous subsection, but with a ten times higher ISM density, i. e. $\rho=10^{-23} {\rm g \ cm}^{-3}$.

\subsubsection{Model \MHH{}}
Assuming a FG cluster with mass $10^6 M_{\odot}$ in a factor 10 denser medium than described in section \ref{M6-Infall24}, the radius of the bubble formed by multiple SN explosions is a factor 3 smaller (see Eq. \ref{eq:Req}). As a consequence, the time required for the cluster to move out of the cavity generated by FG SNe and reach the pristine gas is shorter. By means of equation \ref{eq:tI},  for $t_{\rm I}=3$ Myr we obtain 
the same value as the \MLL{} model. 

The infall passes through the center of the cluster at 4 Myr, which is also the time at which dense gas starts to be accumulated and star
formation is ignited. 
This is before the onset of AGB ejecta, so He-poor stars begin forming very quickly out of the pristine ISM gas. When the AGB ejecta are injected at $7.7$ Myr, more enriched gas is available and then stars start to form with larger helium abundances.

The third rows in Figures \ref{fig:dens} and \ref{fig:temp} illustrate the time evolution in the simulation. At $10$ Myr a dense tail of cold gas has already formed. The newly-formed SG stars are distributed across a wide region whose shape reflects the relative motion between the cluster and the ISM. At $29$ Myr, an elongated dense central stellar component is visible, whereas the new stellar population is settling into a more regular,
nearly ellipsoidal distribution. This process continues towards later times and eventually SG stars can be seen almost everywhere in the box, while the central dense region slightly expands along the tail. The final stellar distribution is significantly more extended than in the case with lower ISM density, i.e. \MHL{}. This can better be seen in in the third-row panels of Fig. \ref{fig:profil}, showing the SG density radial profile computed at different times. At first, in the innermost $\sim 4$ pc, the dominating stellar population has an intermediate helium abundance, whereas stars belonging to the lowest $Y$ bins are present in significant quantities. Almost no He-rich stars (with $Y>0.33$) are present, due to the dominant fraction of ISM gas fuelling the star formation.  At distances $>10~pc$ from the centre, the SG population becomes decreasingly He-enriched. Everywhere in the cluster, the stellar density is still dominated by FG stars. As time progresses, the SG starts to dominate the stellar density in the inner $1$ pc and the SG population becomes decreasingly He-enriched, i.e. increasingly fuelled by the pristine ISM gas, and much less enriched than in the previous models. 
The SG is more compact than the FG of stars, but it is significantly less compact here than in the other models. The central density here at $71$ Myrs is $\gtrsim 10^{-17}$ g cm$^{-3}$ ($\gtrsim 1.5 \times 10^5$ $M_{\odot} \ {\rm pc}^{-3}$), about a factor ten lower than the \MHL{} model.

The lower left panel of Fig. \ref{fig:dist} shows that the stellar distribution in $Y$ is totally dominated by ISM gas ($Y = 0.246$), though it has a flat tail of enriched helium that peaks at the opposite end of $Y \approx 0.32$. This evolves with time towards a slight but rather broad He enrichment due to dilution of AGB ejecta by the pristine gas, peaking at $ Y\approx 0.265$ at the final time.

The final SG stellar mass formed in this model is $2~\times~10^5$ M$_{\odot}$ and the final half-mass radius of SG stars is 1.5 pc.

\subsubsection{Model \MLH{}} 
This model presents the physical conditions in which the early SN feedback of massive FG stars is the least effective, due to the low mass of the FG population and dense ISM gas it has to fight against. 
As a consequence, this is the model with the earliest infall.  

The time-evolution of this run is shown in the bottom rows of Figures \ref{fig:dens} and \ref{fig:temp}). Due to the low FG stellar mass, the central accumulation of mass and star formation are significantly delayed with respect to the other simulations.
The cluster accumulates matter at its centre rather slowly and starts to form stars at about $30$ Myr. 
A significant enhancement of the central density of the gas becomes visible at this time (central panel at the bottom of Figure \ref{fig:dens}), along with a compact, newly born stellar component. The temperature map highlights the appearance of a tenuous cold tail downstream of the centre (Figure \ref{fig:temp}), through which the cluster starts to slowly accrete gas. As this process continues, the cluster is able to form more SG stars.

The bottom-row panels of Fig.\ref{fig:profil} show that the emerging SG population grows slowly, is compact, and dominated by  intermediate He abundances. There is a rather strong contrast with the low-mass low-density model (\MLL{}; second row), which has a much more enriched SG population, due to the clusters ability to hold on to its AGB ejecta. 

In Fig. \ref{fig:dist}, we can see that the emerging stellar component is 
mildly enriched with He as due to a composition made of an approximately equal mixture of pristine gas and AGB ejecta and,
as in te other runs, it evolves towards a broader He distribution. 

Eventually, this model produces the smallest final SG mass (2.3 $\times~10^3$ M$_\odot$) and a final half-mass radius of $0.14$ pc. The mass fractions of stars born out of pristine gas and AGB ejecta are nearly equal (Table \ref{tabl2}). 

In summary, our high-density simulations present two different kinds of results. In the case of the higher mass cluster, more pristine matter can be accreted and the SG stellar mass is much higher than the case with more diffuse infall. SG stars are more extended with a denser ISM and the have lower He abundances. In general, this case gives results similar to the simulation with a very massive cluster (C19). The case of the lower mass cluster is quite different.  Not only very few stars form, but the SG stars are still concentrated in a radius of less than 1 parsec. The SG to FG mass is similar for the two masses in the diffuse case (Table \ref{tabl2}), but very different in the dense ISM case. In the low-mass cluster case, the results are fairly similar regardless of the ISM density.

\subsection{Second generation star formation}
\begin{figure}
   \centering
   \includegraphics[width=1.\linewidth]{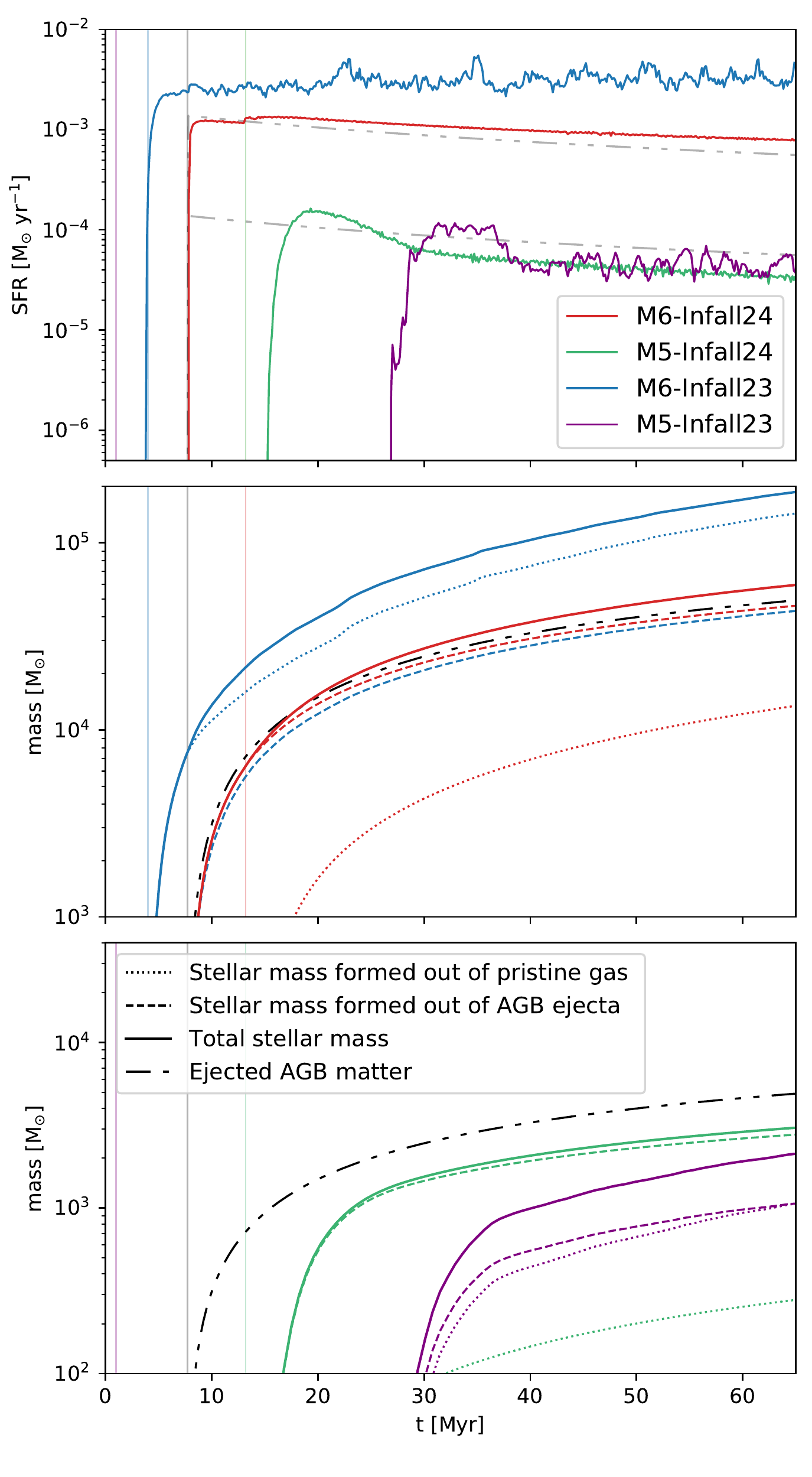}
   \caption{\textit{Top panel}: SG star formation rates vs. time for our simulated clusters. \textit{Middle and lower panels}: cumulative SG stellar mass formed in the $10^6~M_{\odot}$ and $10^5~M_{\odot}$ clusters, respectively. The curves are colour-coded as indicated in the legend in the bottom panel. The solid, dashed and dotted lines show the total SG stellar mass, the mass formed from AGB ejecta, and from pristine gas, respectively. 
The coloured and grey vertical lines mark the times at which the infalling gas crosses the center of the cluster in each model 
  and the onset of the injection of 
 AGB ejecta, respectively.}
  \label{fig:sfr}
\end{figure}

The top panel of Fig. \ref{fig:sfr} shows the star formation rates of our simulations.
Star formation in the  $10^6~M_{\odot}$ models starts when the accreted gas reaches the center (high-density model) or the AGB ejecta initiate (low-density model). This is different in the $10^5~M_{\odot}$ cases. Here, star formation begins with a time delay compared to the times of the onset of AGB and the time at
which the infall crosses the cluster centre. This can be attributed to the weaker gravitational potential of these clusters and the smaller amount of ejected AGB gas. The time delay is longer in the case of \MLH{}, due to the higher ram pressure from the ISM. 

In Fig. \ref{fig:sfr}, we also show the rate of mass injection from AGB stars with dash-dotted lines. The more massive cluster runs settle to star formation rates larger than the AGB mass injection rate, whereas the lower mass runs settle to SFRs lower than the injection rate (and approximately the same for both infall densities). This is because the more massive cluster is capable of retaining the AGB ejecta both at low and high density, but the amount of gas accretion is clearly larger if the accreting ISM density is larger. In \MHL, the amount of mass that can be retained by the cluster and transformed into stars is dominated by the injection of AGB ejecta, and hence the SFR is regulated mostly by the rate of stellar mass return. On the other hand, the lower mass cases show a much lower capability to retain both the AGB ejecta and the pristine gas, especially at the start of the simulations. As a result, this cluster exhibits lower SFRs than the rate of stellar mass return.  

The middle and bottom panels of Fig. \ref{fig:sfr} show the cumulative mass of SG stars (solid lines) formed as a function of time in our $10^6~M_{\odot}$ and $10^5~M_{\odot}$ simulations, respectively. In these plots, we also show how the SG mass is divided into pure AGB ejecta (dashed lines) and pristine gas (dotted lines). For reference, the cumulative AGB ejecta (computed from equation \ref{AGB_ej}) are shown as dash-dotted lines. The most massive SG component ($\sim 2 \times 10^5~M_{\odot}$) is obtained in the \MHH{} model, which is mostly formed out of pristine gas.  In this model, star formation starts before FG stars enter their AGB phase. Here AGB ejecta compose about $25$ percent of the SG mass fraction.

In the \MHL{} model this is reversed, as at all times the majority of gas forming the SG stars are AGB ejecta, composing about $80$ percent of the SG stellar mass at the final time (and more at earlier times).  Another notable feature of both the \MHH{} and \MHL{} simulations is that the AGB ejecta are retained and converted into stars very efficiently (as shown by the close vicinity of the black dash-dotted and red and blue dotted lines in the upper panel of Fig. \ref{fig:sfr}). 

The $10^5~M_{\odot}$ cluster shows a much lower capability to retain mass and to form new stars. In the \MLL{} model, star formation starts a few Myr after the infall and, at all times, it is almost completely dominated by AGB ejecta. The higher-density \MLH{} run sees a later beginning of the SF phase, occurring approximately at 27 Myr. In this model the fractions of stars formed out of AGB ejecta and pristine gas are comparable at later times, whereas the former dominates at earlier times. In both cases, the cumulative AGB mass is significantly larger than the stellar mass, with a larger retained fraction in the lower density model. One interesting aspect of these models is the larger amount of stellar mass formed at a low ISM density, the reverse of the results of the $10^6~M_{\odot}$ runs.  This further outlines how the capability to form stars is sensitive to both accretion and ram pressure from the external ISM.

 
 
%

\section{Discussion}\label{discussion.sec}
\begin{figure*}
\centering
\includegraphics[width=\linewidth]{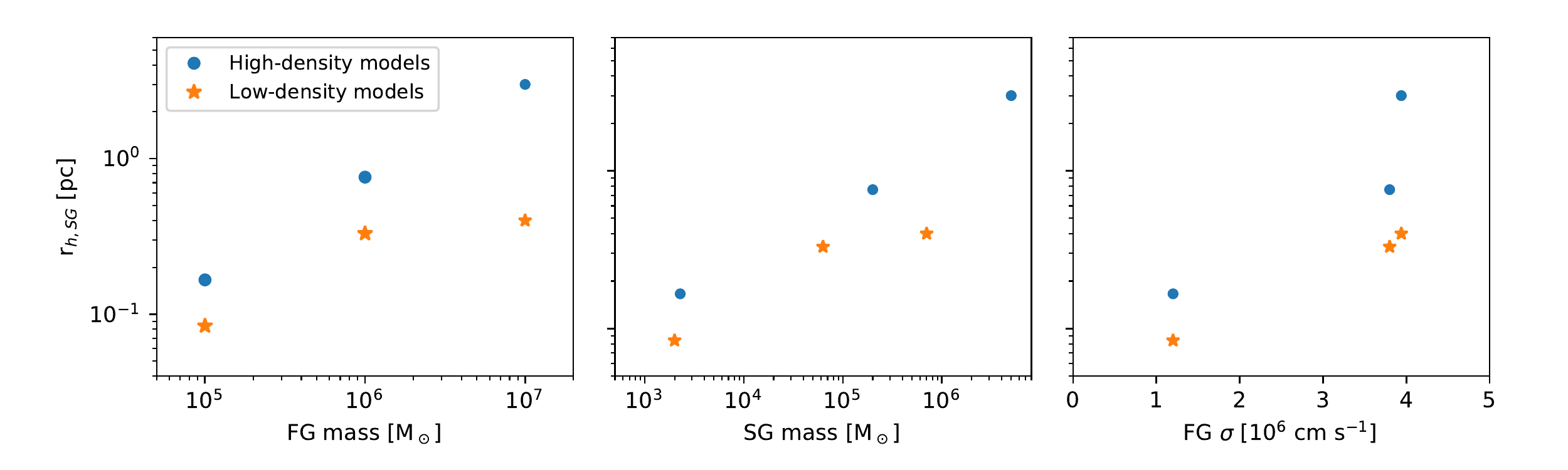}
\caption{Correlations between half-mass radius of SG stars and FG cluster mass (left panel), SG cluster mass (middle panel) and velocity dispersion of FG stars (right  panel) computed at the final time of our simulations (71 Myr). The orange stars and solid blue circles show the results obtained in our low-density and high-density models, respectively. The data shown for a FG mass of $10^7 M_{\odot}$ are from C19. Note that the half-mass radius of the FG clusters of masses of $10^5$ and $10^6 \Msun$ are set to be 4pc in our simulations, whereas it is adopted to be 30 pc in the $10^7 \Msun$ cluster shown here and taken from C19. Depending on the density of the pristine gas, positive correlations can be seen in all panels.
} 
\label{fig:rhm}
\end{figure*}

Our hydrodynamical simulations show how clusters with FG masses of $10^5~M_{\odot}$ and $10^6~M_{\odot}$ moving through diffuse gas can accrete enough mass to form a new stellar generation. This result has been found assuming two different, realistic values for the external ISM density, i. e. $\rho=10^{-24} \ {\rm g~cm}^{-3}$ and $\rho=10^{-23} \ {\rm g~cm}^{-3}$ or particle densities $\approx 1$ and 10 ${\rm cm^{-3}}$, typical for normal main-sequence star-forming galaxies at low and high redshifts \citep{federrath2017,wardlow2017}.  
After describing the results of our simulations computed for FG components spanning three orders of magnitude in mass, in this Section we 
show how some important structural properties scale with cluster mass. 
Fig.~\ref{fig:rhm} shows three scaling relations for the SG half-mass radius, $r_{\rm h,SG}$, computed from our simulations, as well as a more massive cluster from C19. These include correlations with the FG mass (left panel), the final SG mass (middle panel), and the FG velocity dispersion (right panel). Positive correlations are found in all three cases, with slopes which depend on the density of the pristine gas. The high-density simulations produce steeper slopes in all three plots. This suggests that the half-mass radius depends on both intrinsic and environmental properties and that the interplay between mass and gas density is crucial in determining the structural properties, such as the size of the SG component. 
We note that in all calculated models, $r_{\rm h,SG}$ is less than the FG half mass radius, $ \rm r_{h,FG}$, i.e. the SG stars are more concentrated.  Since the ratio of $ \rm r_{h,SG} / r_{h,FG}$ plays a key role in the mixing of two populations and their dynamical evolution \citep{vesperini2021}, further parametric studies on $ \rm r_{h,FG}$ need to be performed in the future.  

We recall that some of the results shown in Fig.~\ref{fig:rhm} are dependent on the assumed initial conditions. 
In our simulations, for the FG we assume scaling radii supported by local observations, 
i.e. by the flat size-mass relation observed  in a variety of stellar clusters of various types and ages \citep{krumholz2019}. 
It is currently not known if the same relation holds also at high redshift, in physical conditions more similar to the ones in which GCs originated.

A somewhat peculiar behaviour which is common to both cluster masses is that the central density of SG stars achieved in the lower density ISM case is higher than in the higher ISM density case (Fig. \ref{fig:profil}). A similar result was found also by C19 in a $10^7~M_{\odot}$ cluster, hence it seems to be independent of the FG mass. This is due to the larger ram-pressure of the high-density model, in which the mass accumulation in the cluster centre is less efficient than in low-density models. The final result is a more massive cluster, but also more extended. In all cases, the final central density is $\gtrsim 10^{-17}$ g cm$^{-3}$ ($10^5$ $M_{\odot} \ {\rm pc}^{-3}$), consistent with the values observed today in GCs (e. g., \citealt{renzini2015}).

The shape of the final He distribution in Fig. \ref{fig:dist} is found to be very sensitive to the model. A final, double-peaked distribution is obtained only with the $10^6~M_{\odot}$ cluster, as was found for a ten times more massive cluster in C19. The lower mass clusters cover a more limited range of helium enhancements for SG stars, showing that these cases do not present the right conditions for an effective dilution. On the other hand, since the predominant gas inside the cluster in the low-density cases is the gas enriched by AGBs, the He distribution tends towards high helium enhancements. However, in all simulations it gradually becomes broader towards lower helium enhancements over time, due to the dilution of AGB ejecta by the pristine gas. 

In Table \ref{tabl2} we present the main properties of the SG components formed in this paper, namely  the final SG mass,
the SG mass fractions and the He enrichment. 

As for the final SG mass fraction, the interplay between mass and ISM density is found to be complex. 
At fixed stellar mass, for the $10^6~M_{\odot}$ cluster 
the SG fraction increases with the ISM density, but the same is not true for the  $10^5~M_{\odot}$ cluster, which presents a slightly more massive SG in the lower ISM density model. The reason for this behaviour is that, due to the lower ram-pressure,
the M5-Infall24 accumulates mass (mostly AGB ejecta) more easily than the M5-Infall23. A more detailed interpretation of the 
relative amount of SG stars obtained in our simulations as a function of cluster mass will be discussed in Sect.~\ref{sec_SG_frac},
where our results will be compared with previous, observationally-based estimates. 


\subsection{SG fraction as a function of the initial mass}\label{sec:sgmf}
\label{sec_SG_frac}
One of the most important observational discoveries in the field of multiple stellar generations in globular clusters is the strong correlation between the number ratio of SG to FG stars and the present-day cluster mass \citep{bastian2018}. We now investigate this correlation in our simulations and assess whether it agrees with observations. 

\cite{milone2017} and \cite{milone2020} studied the correlation between the SG to FG number ratio and cluster mass in globular clusters with various structural parameters. In their study, they used archival observational data, such as the extant GC comprehensive catalogues of \citet[][2010 edition]{harris1996} and \citet{mclaughlin2005}. 
\citet{milone2017} analysed high-precision HST photometry along the RGB for 57 Galactic GCs and showed that the SG to FG number ratio in their sample correlates with the absolute luminosity and the mass of the GC. In \citet{milone2020} these studies were extended to consider also various multiple-generation GCs belonging to both Magellanic Clouds (MC), in order to assess whether this correlation is dependent on the properties of the host galaxy. They conclude that a strong correlation between the present-day number ratio of SG to FG and the present-day cluster mass is found for all GCs with MPs. 

The main dynamical effect of long-term cluster evolution is mass loss. In this process,
due to stellar evolution and other dynamical effects such as two-body relaxation, tidal stripping and shocks 
the stars that are lost from clusters add to the population of field stars (e. g., \citealt{Heggie2003, lamers10}).
The final result of this process can be a significant decrease of the mass of the cluster; as a consequence,  
it is therefore expected that GCs were initially more massive than now. 
Recently, \citet{baumgardt2018} computed initial cluster masses for a comprehensive sample of Milky-Way GCs. 
Based on these results, \citet{milone2020} showed that a correlation exists between the present-day SG-to-total number ratio $N_{\rm SG}/N_{\rm t}$, where $N_{\rm t}={ N_{\rm FG}+N_{\rm SG}}$ and the initial cluster mass. 

\begin{figure}
\centering
\includegraphics[width=\linewidth]{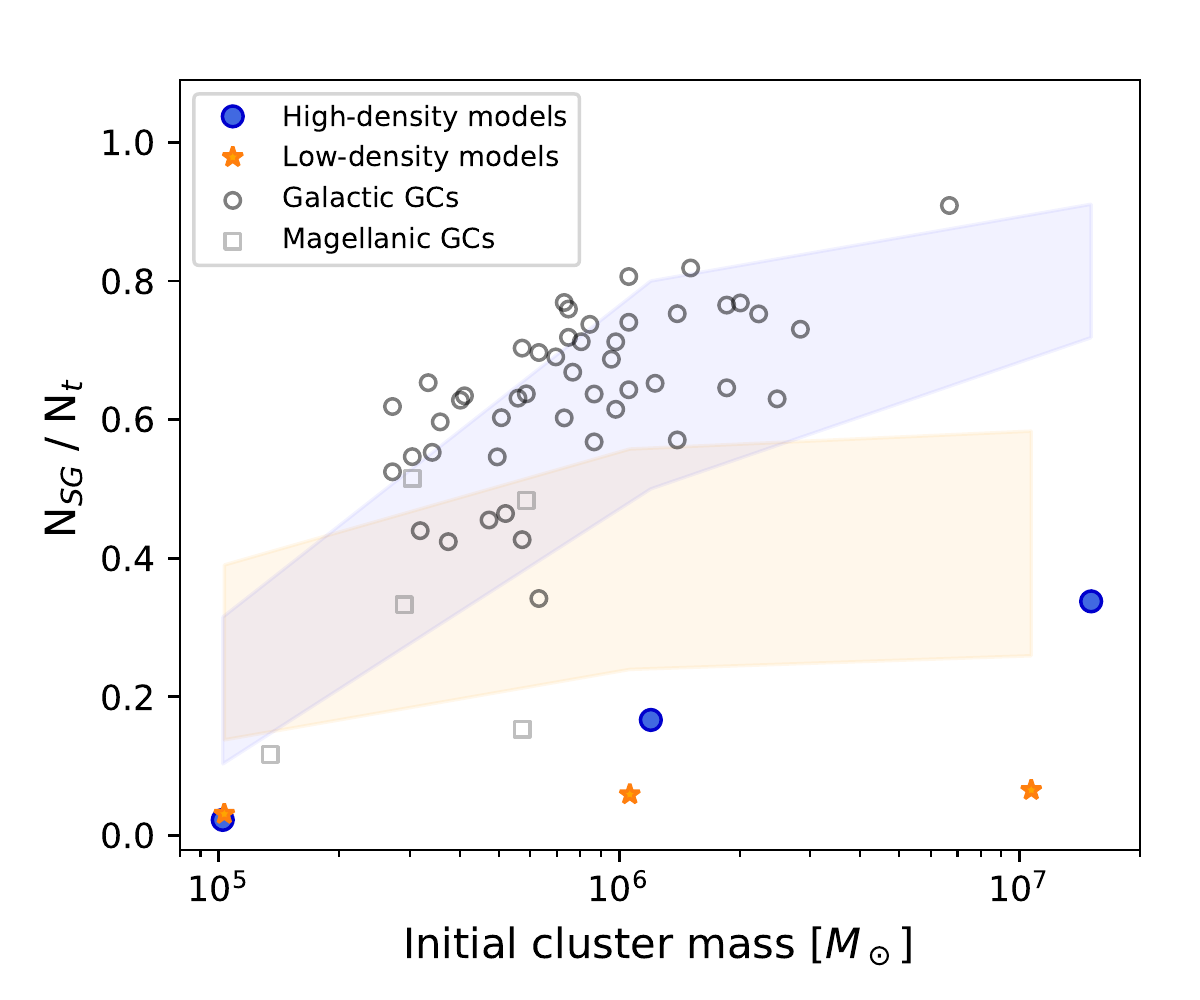}
\caption{SG-to-total number ratio ($N_{\rm SG}/N_{\rm t}$) as a function of cluster mass computed at the final time of our simulations (71 Myr), compared to an observational compilation of \citet{milone2020} of Galactic and Magellanic GCs. The orange stars and solid blue circles show the results obtained in our low-density and high-density models, respectively. The simulation results for a FG mass of $10^7 M_{\odot}$ are from C19.  The open grey circles and squares are number ratios observed in present-day MW and MC clusters, respectively. The blue (high-density models) and the orange shaded regions (low-density models) show the areas covered by the SG/FG number ratios corrected by factors between 5 and 20 to account for the effects of long-term dynamical evolution.
(see Sect.~\ref{sec:sgmf} for details).
}
\label{fig:frac}
\end{figure}

In Figure \ref{fig:frac}, the open circles and squares show the
observed correlation between the present-day SG-to-total number ratio,
$N_{\rm SG}/N_{\rm t}$ and initial cluster mass for Galactic and
Magellanic GCs. We also show the results of our high-density (blue
circles) and low-density (orange stars) models obtained at the final
simulation time of 71 Myr\footnote{To compare our results with the observed SG fractions, for each model a calculation of the \textit{number}
ratios between SG and FG stars is required. 
The derivation of these quantities from the mass fractions presented in Table 2 requires an assumption regarding the stellar IMF. 
For this purpose, for the sake of simplicity we assume that FG and SG stars have the same IMF. 
However, we have ascertained that the assumption of different IMFs for FG and SG stars does not have a significant impact on the results discussed in 
Sec.~\ref{sec_SG_frac}}. 
Both the simulations and the observed data
show a positive correlation between the SG-to-total number ratio and
cluster mass. 
The different slope in the models is mostly related to the relative 
amount of SG stars formed out of the AGB ejecta. In the low-density
models, the SG stellar mass is dominated by the cumulative AGB ejecta
which, to a first approximation, is proportional to the FG mass (Eq. \ref{AGB_ej}), 
therefore the resulting $N_{SG}/N_t$ ratio results weakly dependent on this
quantity.  On the other hand, in the high-density models with mass
$\ge 10^6 M_{\odot}$ the total SG mass is dominated by the pristine gas
(Fig. \ref{fig:sfr}) and grows with the cluster mass but its dependence on the FG mass is not linear. 
A qualitative explanation may be found in the analytic formula for the
Bondi-Hoyle-Lyttleton accretion rate, namely $\dot{M} \propto G M^2 \rho v^{-3}$, where
$M$ is the mass of the accretor, $\rho$ is the ambient density and $v$ the relative velocity between the accretor and the ISM
\citep{edgar04}. 
Therefore, if the accreted mass is dominated by the amount accreted from the pristine gas, to a first approximation 
the SG mass is proportional to $M_{FG}^2$, which qualitatively accounts for the steeper increase of $N_{SG}/N_t$  as a function of mass. 

However, the observed $N_{\rm SG}/N_{\rm t}$ ratios are significantly higher than in our simulations. 
It is important to bear in mind that a direct comparison of our results with the observed present-day properties of GCs 
require taking into account the long-term dynamical evolution of the clusters.


The long-term dynamical evolution and the preferential depletion of FG stars, which are characterised by a more diffuse distribution
than SG stars (see Fig.~\ref{fig:profil}) is expected to lead to a progressive increase of the $N_{\rm SG}/N_{\rm t}$ ratio. 
We attempt to approximately account for such an increase by means of an alternative, simplified approach, which consists in 
rescaling the initial SG to FG ratios by factors between 5 and 20 \citep{renzini2015}. 
Such factors are commonly regarded as indicative lower and upper limits for the ratio between the present-day and initial mass
of GCs required to address the so-called 'Mass Budget' problem. In this framework, GCs had to be initially much massive than today
in order to deliver significant amounts of matter with the composition needed to explain the observed chemical anomalies 
(\citealt{D'Ercole2008,renzini2015,bastian2018}, C19). 
This rescaling gives us an approximate range for the present-day value of the $N_{\rm SG}/N_{\rm t}$ ratio of our models. 
The rescaled $N_{\rm SG}/N_{\rm t}$ values are shown with blue and orange shaded regions in Fig. \ref{fig:frac} for the high-density and low-density models,
respectively. Incorporating this correction  produces good agreement between the high-density models and the observations. 
The fact that the slope of the observed relation is roughly accounted for is a further confirmation of the key role of the pristine gas accretion in the
formation of SG stars, which is clearly of minor importance in the low-density models. 

It remains important in the future to model the long-term dynamical evolution of the systems emerging from our simulations. The use of an N-body approach and a detailed modelling of the external tidal field of the host galaxy will be useful to investigate the evolution of the $N_{\rm SG}/N_{\rm t}$ ratio, as well as the dynamical evolution of the various SG sub-populations characterised by different chemical abundances.

\subsection{Second generation He enrichment versus cluster mass}
Recent results from HST photometry \citep{milone2015,milone2017,lagioia2019,milone2020} reveal a correlation between the observed He enhancement and initial cluster mass in GCs hosting MPs, such that more massive clusters have larger He enhancements.
\begin{figure}
  \centering
  \includegraphics[width=\linewidth]{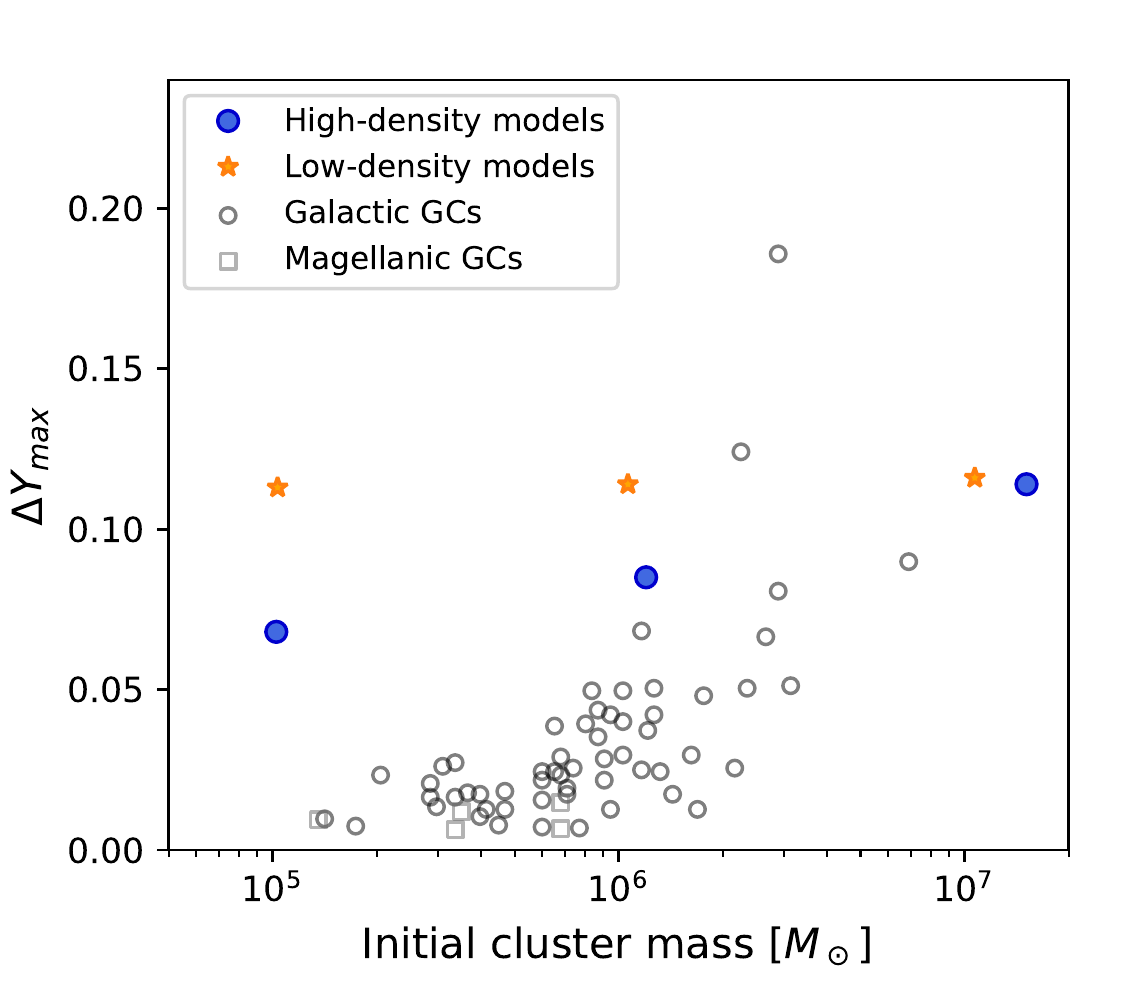}
  \caption{Maximum helium enhancement in SG stars as a function of cluster mass. The orange stars and solid blue circles show the results obtained in our low-density and high-density simulations, respectively. The simulation results for a FG mass of $10^7 M_{\odot}$ are from C19. The open grey circles and squares are the helium enhancements observed in present-day MW and MC clusters, respectively.} \label{fig:DY}
\end{figure}


In Figure \ref{fig:DY} we show the maximum He enhancement, defined as $\DeltaYmax=Y_{\rm max}-Y_0$ (where 
$Y_{\rm max}$ and $Y_0$ are the maximum and the FG He mass fraction, respectively) as a function of cluster mass in our high-density
(filled blue circles) and low-density (filled orange stars) models, including also the models of $10^7 \Msun$ FG mass of C19.
This figure shows that the chemical composition of SG stars can vary with both the FG mass and density of the external ISM.

We also show observational results assembled by \citet{milone2020} for MW and Magellanic GCs with open grey circles and squares, respectively.

Observationally, $\DeltaYmax$ represents the difference in helium abundance between SG and FG stars and was computed by means of the method introduced
by \citet{milone2013} (see also \citealt{lagioia2019}). In brief, the difference of the He abundance in stars of different generations
is measured from photometric data, i. e. by means of color-magnitude diagrams in various HST filters covering a wide range of wavelengths 
and from the estimated color difference between separated sequences.  
By comparing the observed colors with those obtained from synthetic spectra, estimates of the relative He abundances can be derived.  

In the observational dataset, the maximum He enhancement is found to increase as a function of the FG mass,  to lie in the range $\DeltaYmax \approx 0-0.2$
and to follow a similarly increasing trend in both Galactic and Magellanic GCs \citet{milone2020}.

The low-density simulations show a $\DeltaYmax$ independent of cluster
mass, whereas a positive correlation is found in the case of the
high-density runs. In this case, even if the predicted relation is shallower than
the observations, this is a promising result. To our knowledge, it has
not previously been attempted to reproduce this relation within a any
scenario for MP formation. 


In the low-density models, the accretion of pristine gas and dilution play a marginal role, therefore the maximum enrichment level
is dictated mostly by the difference between the abundance of the most He-enriched stars and the FG, a factor which is independent from mass. 
In fact, as seen in Fig.~\ref{fig:dist}, these models contain stars formed out of almost pure AGB ejecta.

In the high-density models, more massive clusters are characterised by more He-enhanced SG stars. 
As visible in Fig.~\ref{fig:profil}, the infall prevents the birth of a substantial population of very He-rich stars in the $10^5~M_{\odot}$ and $10^6~M_{\odot}$ clusters. 
This does not occur in the $10^7~M_{\odot}$ model, in which dilution is delayed with respect to the lower mass models 
and which is more efficient in retaining and transforming into new stars the ejecta of the most massive AGBs (C19). 
This model presents a $\DeltaYmax$ value very similar to the one of its homologous, lower-density model. 
 
In order to better match the observed slope, a stronger dilution of stellar winds is required in lower mass clusters, perhaps achievable
by considering an even denser ISM, or by preventing the retention of the most He-enriched ejecta in lower mass clusters by means of feedback processes such as ionising radiation
(\citealt{gavagnin2017,chantereau2020}). 
The inclusion of such processes is currently in progress and their effects on the He abundance of SG stars will be presented in a future paper.

\subsection{Caveats}
The present study makes a few simplifications. In this Section, we will discuss the most important ones, their implications and
how they  could be possibly investigated further in the future.

First, we model the FG of stars by means of a static Plummer density
profile. This might lead us to omit a few dynamical effects.  
In principle, the accretion of mass and subsequent star formation might lead to further 
contraction of the FG stellar distribution, which is not likely to 
affect the star formation history but it could have effects on its 
long-term dynamical evolution. The study of the subsequent evolution 
requires methods different than the ones discussed in 
the present paper, i.e. an N-body approach to model the evolution of 
the clusters in an external tidal field (e. g., \citealt{mastrobuono19}).
Work is already in progress to address these topics and  
will be the subject of a future paper.

Moreover, we assume a constant value of 20 km s$^{-1}$ for the
velocity of the pristine gas relative to the GC, of the order of the isothermal sound speed of a medium characterised by a  temperature of
$10^4$ (\citealt{D'Ercole2016}, C19).
The adoption of a different velocity might affect the timescale over which the clusters starts accreting pristine gas,
i.e. the time of the infall, and, if too large, even its capability to accrete mass. 
In principle, a different velocity could delay or anticipate the infall time but it 
would not prevent gas accretion as long as the relation $\sigma^2>{V^2_{p g}+c^2_s}$ is satisfied \citep{lin2007}.

It is expected that idealised, core potentials moving at high velocities have
density enhancements and characteristic radii that decrease with increasing velocity \citep{naiman2011}. 
However, to estimate even approximately the effects of a different velocity on gas accretion in our model is not feasible. 
In fact, a detailed analytic formalism for the accretion of gas onto core potentials in realistic conditions does not exist,
since real systems include physical processes (such as radiative cooling and star formation) that cannot be studied without a numerical approach.

The results of the lower mass models show that even if at the beginning the above relation is not satisfied, 
the density field is expected to be perturbed by the presence of the cluster and even a small density enhancement in the centre might lead
to a local, slight increase of the velocity dispersion, which might eventually result in further accretion and ignite star formation.

For the sake of simplicity, we consider only the feedback of FG AGB
stars and neglect the injection of energy by other sources,
such as SNe or other forms of feedback, e. g. ionising radiation. Our simulations are 
  stopped at $\approx 100$ Myr after the formation of the cluster,
  corresponding to the onset of FG type Ia SNe.  However, results
  based on 3-D simulations have shown that in realistic conditions,
stellar feedback from sparse energy sources is not very efficient in
dispersing the cold star-forming gas \citep[][]{romano2019}. 
\cite{Lacchin21} studied how the explosions of Type Ia SNe affect the star formation
history and the chemical properties of SG stars in massive GCs of $10^7~M_{\odot}$.
They used an initial setup similar to ours, i.e. a cluster in motion with respect to a uniform gas distribution  and tested
different assumptions for the gas density and for the type Ia SN delay time distribution.
Their results indicate that type Ia SNe are able to severely limit SF only assuming an ambient gas density of $\rho = 10^{-24}$ g cm$^{-3}$,
whereas their higher-density case ( $\rho = 10^{-23}$ g cm$^{-3}$) is weakly affected by SN explosions, with a final SG mass similar
to the one obtained without SNe Ia.
In the future, it will be very important to extend the study of \cite{Lacchin21} to cluster of different masses, like the ones considered
in this paper, in order to improve our  still limited understanding of the effects of discrete feedback sources on the ISM and, most of all, 
to probe the effects of different numbers of SNe explosions on mass accretion and on star formation in young GCs.

  \subsection{On the requirement of a truncated IMF and on SG massive stars}
  \label{sec_IMF}
The adoption of a IMF truncated below $8-10~M_{\odot}$ is probably the strongest assumption required by the AGB scenario. 
This assumption is necessary in oder to avoid that the SG is significantly enriched with Fe with respect to
FG stars, and also in order to avoid the suppression of star formation due to stellar feedback.  
In this section, we analyse a few theoretical and indirect arguments to support this assumption.

Current observational studies of the variation of the stellar mass functions in the GCs concern
only a very limited mass range and low stellar masses ($\le 0.75 M_{\odot}$, \citealt{cadelano20}). 
Indirect arguments in support of this assumption can be found 
in the framework of the integrated galactic IMF theory (IGIMF, \citealt{weidner2005}).   
In this picture, stars can form only in clusters and, since available observations indicate that the
cluster mass function is a power law with index $\sim -2$ \citep{zhang1999}, most stars in the Universe form in low-mass
clusters. 
The requirement that the most massive star within a cluster cannot exceed the total
cluster mass implies that low-mass clusters will present a deficiency of massive stars \citep{haas2010}. 
This also implies that if a system is characterised by a particularly low star formation rate, 
in principle a generation of stars deprived of massive stars can originate. 
These arguments can explain the absence of massive stars only in very low mass custers,
such as Taurus or IC348 (\citealt{luhman2004}) . 
A few works carried out in this framework have shown that the complete lack of massive stars (i.e. stars with $M\ge 8-10~M_{\odot}$)
can occur in systems with an average SFR $\le 10^{-4}~M_{\odot}/yr$ (\citealt{calura2010}; \citealt{recchi2014}; \citealt{yan2017}). 

The average star formation rates in our $10^5 M_{\odot}$ cluster are between $\approx 3\times 10^{-5} ~M_{\odot}/yr$ and $4.5\times 10^{-5}~M_{\odot}/yr$,
hence below this value.
However, in the case of the $10^6 M_{\odot}$ cluster, SFR values are typically larger than $10^{-4}~M_{\odot}/yr$, 
hence the complete absence of massive stars cannot be justified within the IGIMF theory. However,
according to the same theory, in low SF regimes the number of massive stars is expected to be considerably reduced with respect to a standard IMF. 

In a recent theoretical study based on hydrodynamic simulations, \citep{bekki2019} have shown that in a dense stellar environment,
the dynamical interaction between the gravitational potential of the globular cluster and collapsing star-forming gaseous clouds can suppress the formation
of massive stars, leading to a truncated IMF for SG stars. The validity of this result requires confirmation in a broader context and needs to be tested further, perhaps via
a comprehensive study of the initial conditions and of the role of crucial parameters such as custer mass, size and 
of processes such as turbulence and proto-stellar feedback. 

Another possible mechanism to avoid the effects of massive stellar feedback in dense
cluster cores involves the dynamical decoupling of a small population of segregated massive stars, which might result
in the eventual expulsion of some of them. 
If a few massive stars form in a dense environment, two-body relaxation is expected to be 
very fast and segregation might occurr well before they can explode as SNe (\citealt{gurkan2004, allison2009}), or
stars can even originate already segregated (\citealt{bonnell2002}). 
Several works have shown that close dynamical encounters in a collisional subsystem formed by
massive stars can result in the ejection of some of them (\citealt{gualandris2004,oh2015,wang2019}).

On the observational side, isolated, runaway massive stars have been frequently observed near young massive clusters, with typical
velocities in excess of $30$ km/s (e. g., \citealt{gies1986,stone1991}, see \citealt{andersson2020} and references therein), 
hence higher than the velocity dispersion values of the systems modelled here.

Testing the possibility that a significant fraction of massive stars might be expelled from systems with features similar
to the ones modelled here requires collisional N-body simulations, which currently have been performed
only for systems much less massive that $10^5$ M$_\odot$ (see \citealt{oh2016}).
Performing this kind of simulations for large numer of particles is notorously challenging \citep{rodriguez2018}; however, other less computationally
demanding techniques exist to model collisional systems, such as  Monte Carlo methods \citep{sollima2021, vesperini2021}.
A model involving these techniques and adopting the outputs of our simulations as initial conditions is already under development. 

Facing the problem of massive stars in the SG is the next step of the roadmap to a more complete theoretical understanding of the origin of MPs within GCs.
This topic demands attention and represents an interesting subject for future work. 

\section{Summary and Conclusions} \label{conclusions.sec}
By means of hydrodynamical simulations, we study the capability of clusters of different masses and moving through a uniform gas
distribution to form a new stellar generation. 

We consider FG clusters of mass $10^5$ and $10^6~M_{\odot}$ moving at a constant velocity through a homogeneous gas with densities $10^{-24}$ and $10^{-23}$ g cm$^{-3}$. These simulations are designed to mimic the encounter of a young cluster with a fresh reservoir of gas, e. g. during its orbit through the disc of a host galaxy. The adopted values for the ISM density are meant to be typical for normal star-forming galaxies, whereas the cluster masses are chosen to complement and extend the study started by C19, who considered the same setup but a more massive cluster of $10^7 M_{\odot}$. As in C19, we assume that the SG population does not contain any type II SN progenitor stars (i.e. no stars with mass $>8 M_{\odot}$) and the simulations are stopped 100 Myr after the birth of the cluster, when the first type Ia SNe are expected to explode and halt the formation of stars. 

In addition to star formation, our simulations include mass and energy return from FG AGB stars, radiative cooling and an infall of homogeneous pristine gas with the same composition as the one of FG stars, entering from one side of the box at constant velocity. 

We are able to study how a few important quantities depend on cluster mass  and to model how the chemical properties of SG stars
(expressed by the He mass fraction Y) depend on the intrinsic and environmental quantities analysed in our work.

Our main results can be summarised as follows. 

\begin{enumerate}
\item The capability to accrete mass is determined by the interplay between (a) AGB feedback, through which the FG stars continuously restore gas,
(b) the mass of the cluster and (c) the ram pressure exerted by the external gas. 
Initially, in the high-mass cluster with diffuse infall (\MHL) the cool  AGB ejecta collect into a cooling flow directed towards the centre of the cluster, where the first SG stars are born, showing the highest He abundances ($Y>0.33$).  When the infall starts, the fact that a compact stellar core is already present is key to drive further accretion and the growth of the SG component. Most of the mass is accreted through an ``accretion column'', namely a dense and cold trail of gas developed after the infalling gas has crossed the cluster centre. 
Soon after the start of the infall, SG stars with intermediate He content between the one of FG stars and the first SG stars
start to appear. In general, stars formed with a lower He fraction are characterised by a more extended distribution than extreme, very He-rich stars.
The final $Y$ distribution function is bimodal and broad, with two peaks at $Y\approx$ 0.3 and 0.36.

\item In the low-mass cluster with diffuse infall (\MLL)
  the mass accretion is initially less efficient due to its shallower gravitational potential. 
In the earliest phases, the ram pressure prevents the cluster to retain a significant amount of matter to ignite star formation,
until a slight density enhancement in the central regions of the cluster is sufficient to increase the accretion rate and to ignite star formation.
At 28.8 Myr a very compact stellar component, mostly composed of He-rich stars, occupies the cluster centre. 
In this case, the infalling gas is weakly perturbed by the presence of the cluster and, at variance with the previous case,
an accretion column is not developed. 
The final $Y$ distribution is flatter and narrower than in the previous model.

The two low-density models have final SG masses between 3.2 $\times 10^3 M_{\odot}$ and 6.3 $\times 10^4 M_{\odot}$,
which correspond to SG-to-FG mass ratios of $\approx$0.03 and 0.06, respectively. 

\item The two high-density models are subject to a stronger ram
  pressure than the lower-density models. The larger ambient density
  in which the higher mass cluster moves leads it to accrete more mass
  and form more stars than in the case with more diffuse infall. This
  is because the cluster starts immediately to accrete gas from the
  infalling ISM through a dense and narrow accretion column, causing
  the formation of a massive and extended stellar component which 
  enhances the gravitational field of
  the cluster. When the mass return from AGB stars begins, their
  ejecta are easily retained by the cluster, collecting in the centre
  and fuelling the formation of a compact stellar component.  The
  early dilution of helium causes the central SG component to be less
  He-rich than in the case with more diffuse infall. The massive high-density model is 
  characterised by a broad final $Y$ distribution, ranging from
  $Y\approx$ 0.25 to 0.33.
  
\item On the other hand, the lower mass cluster forms SG stars less efficiently when subjected to a denser infalling gas. This is the model in which star formation starts at the latest time due to its low FG stellar mass and strong ram pressure. At later times as more mass is accumulated at the centre, a compact central SG stellar component with intermediate He enhancement (Y$\approx 0.3$) forms, along  with a smaller, less concentrated population of stars with semi-pristine abundances. This is the model with the narrowest He distribution at the final time, ranging from $Y\approx 0.27$ to $Y \approx 0.31$.

The two high-density models are characterised by broader range for the final SG masses, i. e. between 2.3 $\times 10^3 M_{\odot}$ and $2 \times 10^5 M_{\odot}$
 corresponding to SG-to-FG mass ratios of $\approx$0.02 and 0.2, respectively. 

\item All the models considered here develop a second stellar generation, with a variety of star formation histories which depend on both the cluster mass and the density of the infalling gas. Matched with previous results presented in C19, our simulations allow us to analyse  cluster-related scaling relations across a dynamical range of two orders of magnitude in mass (from $10^5 M_{\odot}$ to $10^7 M_{\odot}$). Positive correlations are found between the half-mass radius of the SG stellar component and FG cluster mass, SG cluster mass and velocity dispersion of FG stars and,  most importantly, the final SG to total mass ratio as a function of both FG and SG mass, with slopes that depend on the assumed density of the pristine gas.
\end{enumerate}

In all simulations, SG stars dominate the central density profile of the cluster, whereas FG stars dominate the outskirts.
The most He-enriched SG stars are also the most concentrated ones, in agreement with observational studies (e. g., \citealt{simioni2016} and references therein). 

We have compared the $N_{\rm SG}/N_{\rm t}$ ratio vs mass with the observational dataset of \citet{milone2020}, which includes a sample of Galactic and Magellanic GCs containing multiple populations. After performing approximate corrections for long-term dynamical evolution,  
good agreement is found between the observational dataset and the high-density simulations, but this is not true for the low-density simulations.
Therefore, in order to account for the observed correlation between $N_{\rm SG}/N_{\rm t}$ and mass, our results indicate that
the SG needs to be formed in dense environments.  

For the first time, we have compared simulation results to the
observed correlation between maximum He enhancement $\DeltaYmax$ and 
cluster mass. Our low-density models show a flat relation between
$\DeltaYmax$ and cluster mass, whereas a positive correlation is found
in the case of the high-density models.
Although caution is needed in the comparison of our results with observations of real GCs,
this is a further confirmation that 
the pristine gas accreted by the cluster has a vital role in
the AGB scenario in order to explain the properties of GCs. 
Even if the  $\DeltaYmax$ - mass relation predicted 
in the high-density models is shallower than observations, this can be regarded as a promising result.
A fair comparison with observational data will require
a modelling of the long-term dynamical evolution, which will be the subject of future work.
Our models need also to be improved with the inclusion of relevant physical mechanisms which are expected to affect mass accretion and star formation,
such as radiative feedback.  
Our understanding of the formation of multiple stellar populations will be improved  
by simulating the formation of GCs in a more realistic environment, e. g. by means of high-resolution cosmological simulations \citep{kimm2016}.

\section*{Data Availability Statement}
The data that support the findings of this study are available from
the corresponding author upon reasonable request.

\section*{Acknowledgements}
We would like to thank Holger Baumgardt, Antonino Milone and Pouria Khalaj for useful discussions and Léo Michel-Dansac for assistance
with computational facilities. AY is grateful to the
Centre de Recherche Astrophysique de Lyon (CRAL) for hospitality during her visit.
FC acknowledges support from grant PRIN MIUR 2017 - 20173ML3WW 001, from the INAF main-stream (1.05.01.86.31) and from PRIN INAF 1.05.01.85.01. 
We acknowledge support and computational resources from the Common Computing Facility (CCF) of the LABEX Lyon Institute of Origins (ANR-10-LABX-66).




\bibliographystyle{mnras}
\bibliography{example} 

\begin{thebibliography}{}
\makeatletter
\relax
\def\mn@urlcharsother{\let\do\@makeother \do\$\do\&\do\#\do\^\do\_\do\%\do\~}
\def\mn@doi{\begingroup\mn@urlcharsother \@ifnextchar [ {\mn@doi@}
  {\mn@doi@[]}}
\def\mn@doi@[#1]#2{\def\@tempa{#1}\ifx\@tempa\@empty \href
  {http://dx.doi.org/#2} {doi:#2}\else \href {http://dx.doi.org/#2} {#1}\fi
  \endgroup}
\def\mn@eprint#1#2{\mn@eprint@#1:#2::\@nil}
\def\mn@eprint@arXiv#1{\href {http://arxiv.org/abs/#1} {{\tt arXiv:#1}}}
\def\mn@eprint@dblp#1{\href {http://dblp.uni-trier.de/rec/bibtex/#1.xml}
  {dblp:#1}}
\def\mn@eprint@#1:#2:#3:#4\@nil{\def\@tempa {#1}\def\@tempb {#2}\def\@tempc
  {#3}\ifx \@tempc \@empty \let \@tempc \@tempb \let \@tempb \@tempa \fi \ifx
  \@tempb \@empty \def\@tempb {arXiv}\fi \@ifundefined
  {mn@eprint@\@tempb}{\@tempb:\@tempc}{\expandafter \expandafter \csname
  mn@eprint@\@tempb\endcsname \expandafter{\@tempc}}}

\bibitem[\protect\citeauthoryear{{Allison}, {Goodwin}, {Parker}, {de Grijs},
  {Portegies Zwart}  \& {Kouwenhoven}}{{Allison} et~al.}{2009}]{allison2009}
{Allison} R.~J.,  {Goodwin} S.~P.,  {Parker} R.~J.,  {de Grijs} R.,  {Portegies
  Zwart} S.~F.,   {Kouwenhoven} M.~B.~N.,  2009, \apjl, 700, L99

\bibitem[\protect\citeauthoryear{{Andersson}, {Agertz}  \&
  {Renaud}}{{Andersson} et~al.}{2020}]{andersson2020}
{Andersson} E.~P.,  {Agertz} O.,   {Renaud} F.,  2020, \mnras, 494, 3328

\bibitem[\protect\citeauthoryear{Bastian \& Lardo}{Bastian \&
  Lardo}{2018}]{bastian2018}
Bastian N.,  Lardo C.,  2018, \aap, 56, 83

\bibitem[\protect\citeauthoryear{Bastian, Lamers, de Mink, Longmore, Goodwin
  \& Gieles}{Bastian et~al.}{2013}]{bastian2013}
Bastian N.,  Lamers H.,  de Mink S.,  Longmore S.,  Goodwin S.,   Gieles M.,
  2013, \mnras, 436, 2398

\bibitem[\protect\citeauthoryear{Baumgardt \& Hilker}{Baumgardt \&
  Hilker}{2018}]{baumgardt2018}
Baumgardt H.,  Hilker M.,  2018, \mnras, 478, 1520

\bibitem[\protect\citeauthoryear{Baumgardt, Hilker, Sollima  \&
  Bellini}{Baumgardt et~al.}{2019}]{baumgardt2019}
Baumgardt H.,  Hilker M.,  Sollima A.,   Bellini A.,  2019, \mnras, 482, 5138

\bibitem[\protect\citeauthoryear{Bekki}{Bekki}{2019}]{bekki2019}
Bekki K.,  2019, \mnras, 486, 2570

\bibitem[\protect\citeauthoryear{Bondi \& Hoyle}{Bondi \&
  Hoyle}{1944}]{bondi1944}
Bondi H.,  Hoyle F.,  1944, \mnras, 104, 273

\bibitem[\protect\citeauthoryear{{Bonnell} \& {Bate}}{{Bonnell} \&
  {Bate}}{2002}]{bonnell2002}
{Bonnell} I.~A.,  {Bate} M.~R.,  2002, \mnras, 336, 659

\bibitem[\protect\citeauthoryear{{Cadelano}, {Dalessandro}, {Webb},
  {Vesperini}, {Lattanzio}, {Beccari}, {Gomez}  \& {Monaco}}{{Cadelano}
  et~al.}{2020}]{cadelano20}
{Cadelano} M.,  {Dalessandro} E.,  {Webb} J.~J.,  {Vesperini} E.,  {Lattanzio}
  D.,  {Beccari} G.,  {Gomez} M.,   {Monaco} L.,  2020, \mnras, 499, 2390

\bibitem[\protect\citeauthoryear{{Calura}, {Recchi}, {Matteucci}  \&
  {Kroupa}}{{Calura} et~al.}{2010}]{calura2010}
{Calura} F.,  {Recchi} S.,  {Matteucci} F.,   {Kroupa} P.,  2010, \mnras, 406,
  1985

\bibitem[\protect\citeauthoryear{Calura, Few, Romano  \& D’Ercole}{Calura
  et~al.}{2015}]{calura2015}
Calura F.,  Few C.,  Romano D.,   D’Ercole A.,  2015, \apjl, 814, L14

\bibitem[\protect\citeauthoryear{Calura, D’Ercole, Vesperini, Vanzella  \&
  Sollima}{Calura et~al.}{2019}]{calura19}
Calura F.,  D’Ercole A.,  Vesperini E.,  Vanzella E.,   Sollima A.,  2019,
  \mnras, 489, 3269

\bibitem[\protect\citeauthoryear{Carretta et~al.,}{Carretta
  et~al.}{2009}]{carretta2009}
Carretta E. e.~a.,  et~al., 2009, \aap, 505, 117

\bibitem[\protect\citeauthoryear{Carretta, Bragaglia, Gratton, Recio-Blanco,
  Lucatello, D'Orazi  \& Cassisi}{Carretta et~al.}{2010}]{carretta2010}
Carretta E.,  Bragaglia A.,  Gratton R.~G.,  Recio-Blanco A.,  Lucatello S.,
  D'Orazi V.,   Cassisi S.,  2010, \aap, 516, A55

\bibitem[\protect\citeauthoryear{Chantereau, Biernacki, Martig, Bastian,
  Salaris  \& Teyssier}{Chantereau et~al.}{2020}]{chantereau2020}
Chantereau W.,  Biernacki P.,  Martig M.,  Bastian N.,  Salaris M.,   Teyssier
  R.,  2020, \mnras, 493, 1306

\bibitem[\protect\citeauthoryear{Conroy \& Spergel}{Conroy \&
  Spergel}{2010}]{charlie2010}
Conroy C.,  Spergel D.~N.,  2010, \apj, 726, 36

\bibitem[\protect\citeauthoryear{D'Ercole, Vesperini, D'Antona, McMillan  \&
  Recchi}{D'Ercole et~al.}{2008}]{D'Ercole2008}
D'Ercole A.,  Vesperini E.,  D'Antona F.,  McMillan S.~L.,   Recchi S.,  2008,
  \mnras, 391, 825

\bibitem[\protect\citeauthoryear{D'Ercole, D'Antona, Ventura, Vesperini  \&
  McMillan}{D'Ercole et~al.}{2010}]{D'Ercole2010}
D'Ercole A.,  D'Antona F.,  Ventura P.,  Vesperini E.,   McMillan S.~L.,  2010,
  \mnras, 407, 854

\bibitem[\protect\citeauthoryear{D'Ercole, D'Antona  \& Vesperini}{D'Ercole
  et~al.}{2011}]{D'Ercole2011}
D'Ercole A.,  D'Antona F.,   Vesperini E.,  2011, \mnras, 415, 1304

\bibitem[\protect\citeauthoryear{D'Ercole, D'Antona  \& Vesperini}{D'Ercole
  et~al.}{2016}]{D'Ercole2016}
D'Ercole A.,  D'Antona F.,   Vesperini E.,  2016, \mnras, 461, 4088

\bibitem[\protect\citeauthoryear{De~Mink, Pols, Langer  \& Izzard}{De~Mink
  et~al.}{2009}]{Mink2009}
De~Mink S.,  Pols O.,  Langer N.,   Izzard R.,  2009, \aap, 507, L1

\bibitem[\protect\citeauthoryear{Decressin, Meynet, Charbonnel, Prantzos  \&
  Ekstr{\"o}m}{Decressin et~al.}{2007}]{decressin2007}
Decressin T.,  Meynet G.,  Charbonnel C.,  Prantzos N.,   Ekstr{\"o}m S.,
  2007, \aap, 464, 1029

\bibitem[\protect\citeauthoryear{Denissenkov \& Hartwick}{Denissenkov \&
  Hartwick}{2013}]{denissenkov2013}
Denissenkov P.~A.,  Hartwick F.,  2013, \mnras, 437, L21

\bibitem[\protect\citeauthoryear{{Edgar}}{{Edgar}}{2004}]{edgar04}
{Edgar} R.,  2004, \nar, 48, 843

\bibitem[\protect\citeauthoryear{Federrath et~al.,}{Federrath
  et~al.}{2017}]{federrath2017}
Federrath C.,  et~al., 2017, \mnras, 468, 3965

\bibitem[\protect\citeauthoryear{Few, Courty, Gibson, Michel-Dansac  \&
  Calura}{Few et~al.}{2014}]{few2014}
Few C.,  Courty S.,  Gibson B.~K.,  Michel-Dansac L.,   Calura F.,  2014,
  \mnras, 444, 3845

\bibitem[\protect\citeauthoryear{Gavagnin, Bleuler, Rosdahl  \&
  Teyssier}{Gavagnin et~al.}{2017}]{gavagnin2017}
Gavagnin E.,  Bleuler A.,  Rosdahl J.,   Teyssier R.,  2017, \mnras, 472, 4155

\bibitem[\protect\citeauthoryear{{Gies} \& {Bolton}}{{Gies} \&
  {Bolton}}{1986}]{gies1986}
{Gies} D.~R.,  {Bolton} C.~T.,  1986, \apjs, 61, 419

\bibitem[\protect\citeauthoryear{Goodman \& Bekki}{Goodman \&
  Bekki}{2018}]{goodman2018}
Goodman M.,  Bekki K.,  2018, \mnras, 478, 3564

\bibitem[\protect\citeauthoryear{Gratton, Sneden  \& Carretta}{Gratton
  et~al.}{2004}]{gratton2004}
Gratton R.,  Sneden C.,   Carretta E.,  2004, \araa, 42, 385

\bibitem[\protect\citeauthoryear{Gratton et~al.,}{Gratton
  et~al.}{2013}]{gratton2013}
Gratton R.,  et~al., 2013, \aap, 549, A41

\bibitem[\protect\citeauthoryear{{Gualandris}, {Portegies Zwart}  \&
  {Eggleton}}{{Gualandris} et~al.}{2004}]{gualandris2004}
{Gualandris} A.,  {Portegies Zwart} S.,   {Eggleton} P.~P.,  2004, \mnras, 350,
  615

\bibitem[\protect\citeauthoryear{{G{\"u}rkan}, {Freitag}  \&
  {Rasio}}{{G{\"u}rkan} et~al.}{2004}]{gurkan2004}
{G{\"u}rkan} M.~A.,  {Freitag} M.,   {Rasio} F.~A.,  2004, \apj, 604, 632

\bibitem[\protect\citeauthoryear{{Haas} \& {Anders}}{{Haas} \&
  {Anders}}{2010}]{haas2010}
{Haas} M.~R.,  {Anders} P.,  2010, \aap, 512, A79

\bibitem[\protect\citeauthoryear{Haffner et~al.,}{Haffner
  et~al.}{2009}]{haffner2009}
Haffner L.,  et~al., 2009, Rev. Mod. Phys., 81, 969

\bibitem[\protect\citeauthoryear{Harris}{Harris}{1996}]{harris1996}
Harris W.~E.,  1996, \apj, 112, 1487

\bibitem[\protect\citeauthoryear{{Heggie} \& {Hut}}{{Heggie} \&
  {Hut}}{2003}]{Heggie2003}
{Heggie} D.,  {Hut} P.,  2003, {The Gravitational Million-Body Problem: A
  Multidisciplinary Approach to Star Cluster Dynamics}

\bibitem[\protect\citeauthoryear{Kimm, Cen, Rosdahl  \& Sukyoung}{Kimm
  et~al.}{2016}]{kimm2016}
Kimm T.,  Cen R.,  Rosdahl J.,   Sukyoung K.~Y.,  2016, \apj, 823, 52

\bibitem[\protect\citeauthoryear{Krause, Charbonnel, Decressin, Meynet  \&
  Prantzos}{Krause et~al.}{2013}]{krause2013}
Krause M.,  Charbonnel C.,  Decressin T.,  Meynet G.,   Prantzos N.,  2013,
  \aap, 552, A121

\bibitem[\protect\citeauthoryear{Kravtsov \& Gnedin}{Kravtsov \&
  Gnedin}{2005}]{kravtsov2005}
Kravtsov A.~V.,  Gnedin O.~Y.,  2005, \apj, 623, 650

\bibitem[\protect\citeauthoryear{Kroupa}{Kroupa}{2001}]{kroupa2001}
Kroupa P.,  2001, \mnras, 322, 231

\bibitem[\protect\citeauthoryear{Kruijssen}{Kruijssen}{2015}]{kruijssen2015}
Kruijssen J.~D.,  2015, \mnras, 454, 1658

\bibitem[\protect\citeauthoryear{Krumholz, McKee  \& Bland-Hawthorn}{Krumholz
  et~al.}{2019}]{krumholz2019}
Krumholz M.~R.,  McKee C.~F.,   Bland-Hawthorn J.,  2019, \araa, 57, 227

\bibitem[\protect\citeauthoryear{{Lacchin}, {Calura}  \& {Vesperini}}{{Lacchin}
  et~al.}{2021}]{Lacchin21}
{Lacchin} E.,  {Calura} F.,   {Vesperini} E.,  2021, \mnras, in press
  (arXiv:2107.07962)

\bibitem[\protect\citeauthoryear{Lagioia, Milone, Marino  \& Dotter}{Lagioia
  et~al.}{2019}]{lagioia2019}
Lagioia E.~P.,  Milone A.~P.,  Marino A.~F.,   Dotter A.,  2019, \apj, 871, 140

\bibitem[\protect\citeauthoryear{{Lamers}, {Baumgardt}  \& {Gieles}}{{Lamers}
  et~al.}{2010}]{lamers10}
{Lamers} H. J.~G.~L.~M.,  {Baumgardt} H.,   {Gieles} M.,  2010, \mnras, 409,
  305

\bibitem[\protect\citeauthoryear{Larsen, Brodie, Grundahl  \& Strader}{Larsen
  et~al.}{2014}]{larsen2014}
Larsen S.~S.,  Brodie J.~P.,  Grundahl F.,   Strader J.,  2014, \apj, 797, 15

\bibitem[\protect\citeauthoryear{Lin \& Murray}{Lin \& Murray}{2007}]{lin2007}
Lin D.~N.,  Murray S.~D.,  2007, \apj, 661, 779

\bibitem[\protect\citeauthoryear{{Luhman}}{{Luhman}}{2004}]{luhman2004}
{Luhman} K.~L.,  2004, \apj, 617, 1216

\bibitem[\protect\citeauthoryear{Mackey, Nielsen, Ferguson  \&
  Richardson}{Mackey et~al.}{2008}]{mackey2008}
Mackey A.,  Nielsen P.~B.,  Ferguson A.,   Richardson J.,  2008, \apjl, 681,
  L17

\bibitem[\protect\citeauthoryear{Maoz, Mannucci  \& Nelemans}{Maoz
  et~al.}{2014}]{maoz2014}
Maoz D.,  Mannucci F.,   Nelemans G.,  2014, \araa, 52, 107

\bibitem[\protect\citeauthoryear{{Marino} et~al.,}{{Marino}
  et~al.}{2014}]{marino14}
{Marino} A.~F.,  et~al., 2014, \mnras, 437, 1609

\bibitem[\protect\citeauthoryear{{Mastrobuono-Battisti}, {Khoperskov}, {Di
  Matteo}  \& {Haywood}}{{Mastrobuono-Battisti} et~al.}{2019}]{mastrobuono19}
{Mastrobuono-Battisti} A.,  {Khoperskov} S.,  {Di Matteo} P.,   {Haywood} M.,
  2019, \aap, 622, A86

\bibitem[\protect\citeauthoryear{McLaughlin \& van~der Marel}{McLaughlin \&
  van~der Marel}{2005}]{mclaughlin2005}
McLaughlin D.~E.,  van~der Marel R.~P.,  2005, \apjs, 161, 304

\bibitem[\protect\citeauthoryear{Milone}{Milone}{2015}]{milone2015}
Milone A.,  2015, \mnras, 446, 1672

\bibitem[\protect\citeauthoryear{Milone, Bedin, Piotto  \& Anderson}{Milone
  et~al.}{2009}]{milone2009}
Milone A.,  Bedin L.,  Piotto G.,   Anderson J.,  2009, \aap, 497, 755

\bibitem[\protect\citeauthoryear{Milone et~al.,}{Milone
  et~al.}{2013}]{milone2013}
Milone A.,  et~al., 2013, \apj, 767, 120

\bibitem[\protect\citeauthoryear{Milone et~al.,}{Milone
  et~al.}{2017}]{milone2017}
Milone A.,  et~al., 2017, \mnras, 464, 3636

\bibitem[\protect\citeauthoryear{Milone et~al.,}{Milone
  et~al.}{2020}]{milone2020}
Milone A.,  et~al., 2020, \mnras, 491, 515

\bibitem[\protect\citeauthoryear{Minniti, Geisler, Peterson  \& Claria}{Minniti
  et~al.}{1993}]{minniti1993}
Minniti D.,  Geisler D.,  Peterson R.~C.,   Claria J.~J.,  1993, \apj, 413, 548

\bibitem[\protect\citeauthoryear{Mucciarelli, Origlia, Ferraro  \&
  Pancino}{Mucciarelli et~al.}{2009}]{mucciarelli2009}
Mucciarelli A.,  Origlia L.,  Ferraro F.~R.,   Pancino E.,  2009, \apj, 695,
  L134

\bibitem[\protect\citeauthoryear{Naiman, Ramirez-Ruiz  \& Lin}{Naiman
  et~al.}{2011}]{naiman2011}
Naiman J.,  Ramirez-Ruiz E.,   Lin D.~N.,  2011, \apj, 735, 25

\bibitem[\protect\citeauthoryear{Naiman, Ramirez-Ruiz  \& Lin}{Naiman
  et~al.}{2018}]{naiman2018}
Naiman J.,  Ramirez-Ruiz E.,   Lin D.,  2018, \mnras, 478, 2794

\bibitem[\protect\citeauthoryear{Nardiello, Piotto, Milone, Rich, Cassisi,
  Bedin, Bellini  \& Renzini}{Nardiello et~al.}{2019}]{nardiello2019}
Nardiello D.,  Piotto G.,  Milone A.,  Rich R.,  Cassisi S.,  Bedin L.,
  Bellini A.,   Renzini A.,  2019, \mnras, 485, 3076

\bibitem[\protect\citeauthoryear{{Oh} \& {Kroupa}}{{Oh} \&
  {Kroupa}}{2016}]{oh2016}
{Oh} S.,  {Kroupa} P.,  2016, \aap, 590, A107

\bibitem[\protect\citeauthoryear{{Oh}, {Kroupa}  \& {Pflamm-Altenburg}}{{Oh}
  et~al.}{2015}]{oh2015}
{Oh} S.,  {Kroupa} P.,   {Pflamm-Altenburg} J.,  2015, \apj, 805, 92

\bibitem[\protect\citeauthoryear{Pflamm-Altenburg \& Kroupa}{Pflamm-Altenburg
  \& Kroupa}{2009}]{pflamm2009}
Pflamm-Altenburg J.,  Kroupa P.,  2009, \mnras, 397, 488

\bibitem[\protect\citeauthoryear{Plummer}{Plummer}{1911}]{plummer1911}
Plummer H.~C.,  1911, \mnras, 71, 460

\bibitem[\protect\citeauthoryear{Portegies~Zwart, McMillan  \&
  Gieles}{Portegies~Zwart et~al.}{2010}]{portegies2010}
Portegies~Zwart S.~F.,  McMillan S.~L.,   Gieles M.,  2010, \araa, 48, 431

\bibitem[\protect\citeauthoryear{Rasera \& Teyssier}{Rasera \&
  Teyssier}{2006}]{rasera2006}
Rasera Y.,  Teyssier R.,  2006, \aap, 445, 1

\bibitem[\protect\citeauthoryear{{Recchi}, {Calura}, {Gibson}  \&
  {Kroupa}}{{Recchi} et~al.}{2014}]{recchi2014}
{Recchi} S.,  {Calura} F.,  {Gibson} B.~K.,   {Kroupa} P.,  2014, \mnras, 437,
  994

\bibitem[\protect\citeauthoryear{Renzini et~al.,}{Renzini
  et~al.}{2015}]{renzini2015}
Renzini A.,  et~al., 2015, \mnras, 454, 4197

\bibitem[\protect\citeauthoryear{{Rodriguez}, {Pattabiraman}, {Chatterjee},
  {Choudhary}, {Liao}, {Morscher}  \& {Rasio}}{{Rodriguez}
  et~al.}{2018}]{rodriguez2018}
{Rodriguez} C.~L.,  {Pattabiraman} B.,  {Chatterjee} S.,  {Choudhary} A.,
  {Liao} W.-k.,  {Morscher} M.,   {Rasio} F.~A.,  2018, Computational
  Astrophysics and Cosmology, 5, 5

\bibitem[\protect\citeauthoryear{Romano, Calura, D’Ercole  \& Few}{Romano
  et~al.}{2019}]{romano2019}
Romano D.,  Calura F.,  D’Ercole A.,   Few C.~G.,  2019, \aap, 630, A140

\bibitem[\protect\citeauthoryear{{Ryon} et~al.,}{{Ryon}
  et~al.}{2017}]{ryon2017}
{Ryon} J.~E.,  et~al., 2017, \apj, 841, 92

\bibitem[\protect\citeauthoryear{Schmidt}{Schmidt}{1959}]{schmidt1959}
Schmidt M.,  1959, \apj, 129, 243

\bibitem[\protect\citeauthoryear{Shima, Matsuda, Takeda  \& Sawada}{Shima
  et~al.}{1985}]{shima1985}
Shima E.,  Matsuda T.,  Takeda H.,   Sawada K.,  1985, \mnras, 217, 367

\bibitem[\protect\citeauthoryear{{Siess}}{{Siess}}{2010}]{siess2010}
{Siess} L.,  2010, \aap, 512, A10

\bibitem[\protect\citeauthoryear{Simioni, Milone, Bedin, Aparicio, Piotto,
  Vesperini  \& Hong}{Simioni et~al.}{2016}]{simioni2016}
Simioni M.,  Milone A.~P.,  Bedin L.~R.,  Aparicio A.,  Piotto G.,  Vesperini
  E.,   Hong J.,  2016, \mnras, 463, 449

\bibitem[\protect\citeauthoryear{{Sollima}}{{Sollima}}{2021}]{sollima2021}
{Sollima} A.,  2021, \mnras, 502, 1974

\bibitem[\protect\citeauthoryear{Sollima, Ferraro, Bellazzini, Origlia,
  Straniero  \& Pancino}{Sollima et~al.}{2007}]{sollima2007}
Sollima A.,  Ferraro F.,  Bellazzini M.,  Origlia L.,  Straniero O.,   Pancino
  E.,  2007, \apj, 654, 915

\bibitem[\protect\citeauthoryear{{Stone}}{{Stone}}{1991}]{stone1991}
{Stone} R.~C.,  1991, \aj, 102, 333

\bibitem[\protect\citeauthoryear{{Sz{\'e}csi} \& {W{\"u}nsch}}{{Sz{\'e}csi} \&
  {W{\"u}nsch}}{2019}]{szecsi2019}
{Sz{\'e}csi} D.,  {W{\"u}nsch} R.,  2019, \apj, 871, 20

\bibitem[\protect\citeauthoryear{{Tenorio-Tagle}}{{Tenorio-Tagle}}{1996}]{tenoriotagle1996}
{Tenorio-Tagle} G.,  1996, \aj, 111, 1641

\bibitem[\protect\citeauthoryear{{Tenorio-Tagle} \&
  {Mu{\~n}oz-Tu{\~n}{\'o}n}}{{Tenorio-Tagle} \&
  {Mu{\~n}oz-Tu{\~n}{\'o}n}}{1998}]{tenoriotagle1998}
{Tenorio-Tagle} G.,  {Mu{\~n}oz-Tu{\~n}{\'o}n} C.,  1998, \mnras, 293, 299

\bibitem[\protect\citeauthoryear{Teyssier}{Teyssier}{2002}]{teyssier2002}
Teyssier R.,  2002, \aap, 385, 337

\bibitem[\protect\citeauthoryear{Ventura \& D'Antona}{Ventura \&
  D'Antona}{2011}]{ventura2011}
Ventura P.,  D'Antona F.,  2011, \mnras, 410, 2760

\bibitem[\protect\citeauthoryear{Vesperini, McMillan, D'Antona  \&
  D'Ercole}{Vesperini et~al.}{2010}]{vesperini2010}
Vesperini E.,  McMillan S.~L.,  D'Antona F.,   D'Ercole A.,  2010, \apjl, 718,
  L112

\bibitem[\protect\citeauthoryear{Vesperini, Hong, Giersz  \& Hypki}{Vesperini
  et~al.}{2021}]{vesperini2021}
Vesperini E.,  Hong J.,  Giersz M.,   Hypki A.,  2021, \mnras, 502, 4290

\bibitem[\protect\citeauthoryear{{Wang}, {Kroupa}  \& {Jerabkova}}{{Wang}
  et~al.}{2019}]{wang2019}
{Wang} L.,  {Kroupa} P.,   {Jerabkova} T.,  2019, \mnras, 484, 1843

\bibitem[\protect\citeauthoryear{Wardlow et~al.,}{Wardlow
  et~al.}{2017}]{wardlow2017}
Wardlow J.~L.,  et~al., 2017, \apj, 837, 12

\bibitem[\protect\citeauthoryear{Weaver, McCray, Castor, Shapiro  \&
  Moore}{Weaver et~al.}{1977}]{weaver1977}
Weaver R.,  McCray R.,  Castor J.,  Shapiro P.,   Moore R.,  1977, \apj, 218,
  377

\bibitem[\protect\citeauthoryear{{Weidner} \& {Kroupa}}{{Weidner} \&
  {Kroupa}}{2005}]{weidner2005}
{Weidner} C.,  {Kroupa} P.,  2005, \apj, 625, 754

\bibitem[\protect\citeauthoryear{{Yan}, {Jerabkova}  \& {Kroupa}}{{Yan}
  et~al.}{2017}]{yan2017}
{Yan} Z.,  {Jerabkova} T.,   {Kroupa} P.,  2017, \aap, 607, A126

\bibitem[\protect\citeauthoryear{{Zhang} \& {Fall}}{{Zhang} \&
  {Fall}}{1999}]{zhang1999}
{Zhang} Q.,  {Fall} S.~M.,  1999, \apjl, 527, L81

\makeatother
\end{thebibliography}




\appendix
\section{Dependence on resolution}\label{sec_res}

\begin{figure}
  \centering
  \includegraphics[width=1.5\linewidth]{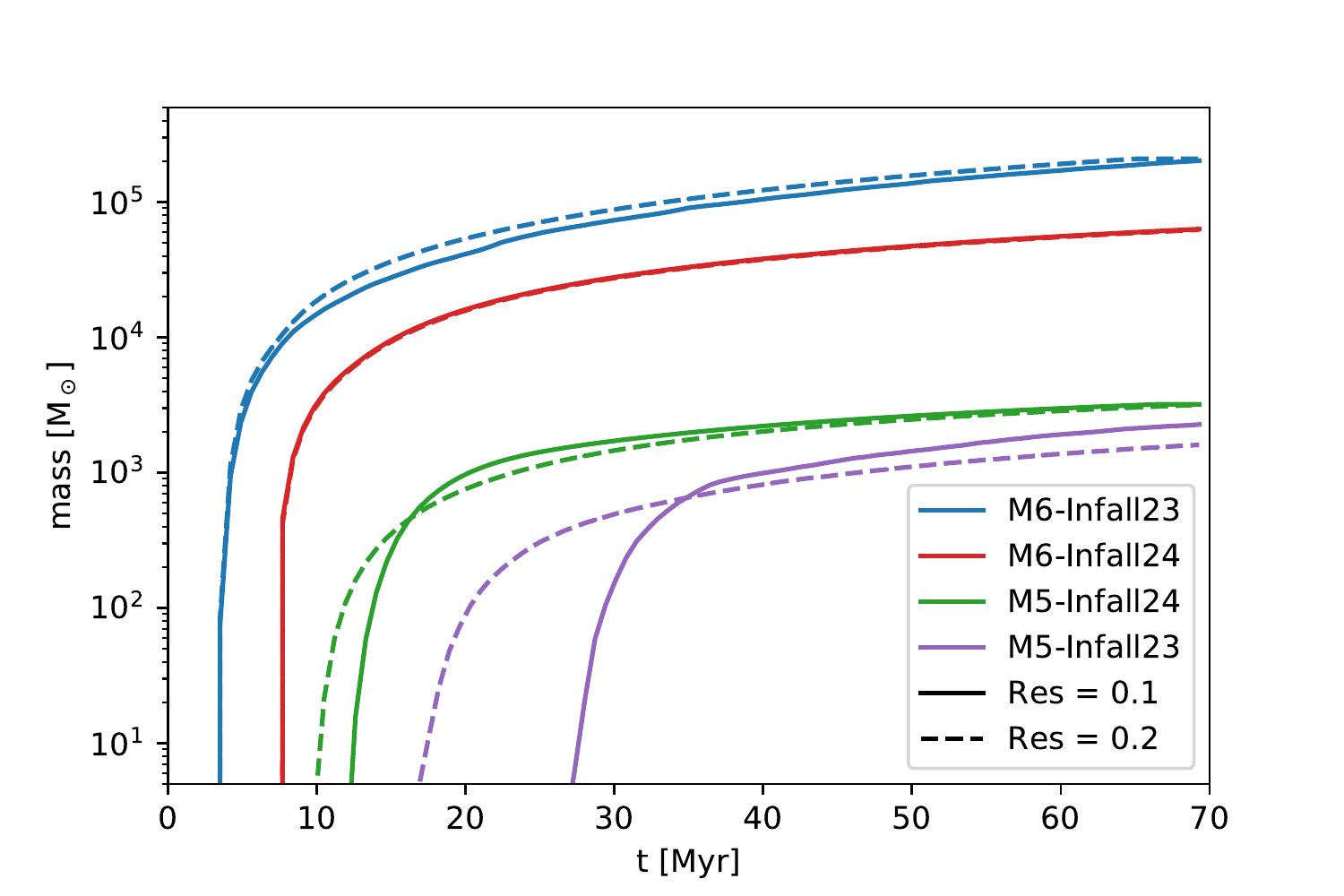}
  \caption{Cumulative SG stellar mass as a function of time formed in the
    $10^5$ and $10^6~M_{\odot}$ model clusters computed at different resolution. The curves are colour-coded as in Fig. \ref{fig:sfr}.
    The solid and dashed lines show the results computed at higher (0.1 pc) and lower (0.2 pc) resolution, respectively } \label{fig:conv}
\end{figure}

The dependence of our results on numerical resolution are presented in Fig. \ref{fig:conv}, where we 
show the time evolution of the cumulative mass for the $10^5$ and $10^6~M_{\odot}$ models in the low-density and high-density gas models computed for different values
of the cell width $\Delta x$.
The two different resolution values at which the simlations have been run  are 0.1 pc (solid lines) and 0.2 pc (dashed lines). 

In both the \MHL~ and \MHH~ models run at different resolution the stellar mass is consistent within a few percent and the final values are significantly similar. 
As for the \MLL~ and \MLH~ models, the runs at different resolution result in a different time at which star formation starts, differing by
less than 10 Myr. This is mostly due to the low amount of accumulated gas, which results in a smaller number of SG stars formed in these models and 
which unavoidably renders their results more subject to numerical noise. 
However, it is worth noting that the final stellar mass values of the  \MLL~ and \MLH~ models computed at different resolution are consistent within a few percent and $\sim$10\%, respectively. 

 \bsp	
\label{lastpage}
\end{document}